\documentclass[a4paper]{article}
 
\usepackage{xspace,colortbl}
\usepackage{amssymb,amsbsy}
\usepackage{graphicx}
\usepackage[T1]{fontenc}
\usepackage[english]{babel}
\usepackage{makeidx}
\makeindex
\usepackage{mathrsfs}
\usepackage{amsfonts,amsthm}
\usepackage[mathscr]{eucal}
\usepackage{cite}
\usepackage{textcomp}
\usepackage{url}
\theoremstyle{definition}

\newtheorem{ese}{Example}[section]

\newtheorem{rem}{Remark}[section]
\theoremstyle{plain}
\newtheorem{prop}{Proposition}[section]
\newtheorem{lemma}{Lemma}[section]
\newtheorem{teor}{Theorem}[section]
\newtheorem{cor}{Corollary}[section]
\begin{document}
\hoffset -2.5cm

\begin{center}
{\LARGE \bf Non-Markovian diffusion equations and processes:\\ analysis and simulations}
\end{center}
\vspace{1cm}
\centerline{\bf Antonio MURA$^1$, Murad S. TAQQU$^2$ and Francesco MAINARDI$^1$}
\vskip .4cm
\begin{center}
{\it   $1.$ Department of Physics, University of Bologna, and INFN,
Via Irnerio 46, I-40126 Bologna, Italy}
\\ URL: http:// www.fracalmo.org
\\
{\it   $2.$ Department of Mathematics, Boston University, Boston, MA 02215, USA}
\\
URL:  http://math.bu.edu/people/murad/
\\
\vskip 0.5truecm
{\bf Revised Version: May 2008}
\\ {\bf in press on Physica A (2008),  doi:10.1016/j.physa.2008.04.035}
\end{center}
\vskip 0.5truecm \noindent
{\bf Abstract}: In this paper we introduce and analyze a class of diffusion type equations related 
to certain non-Markovian stochastic processes. 
We start from the forward drift equation which is made non-local in time by the introduction of a 
suitable chosen memory kernel $K(t)$. 
The resulting non-Markovian equation can be interpreted in a natural way as the evolution equation of 
the marginal density function of a random time process $l(t)$. 
We then consider the subordinated process $Y(t)=X(l(t))$ where $X(t)$ is a Markovian diffusion. 
The corresponding time evolution of the marginal density function of $Y(t)$ 
is governed by a non-Markovian Fokker-Planck equation which involves the memory kernel $K(t)$. 
We develop several applications and derive the exact solutions. 
We consider different stochastic models for the given equations providing path simulations.    

\section{Introduction}\label{s1}    
In this introduction, we describe and motivate the themes developed in the paper. 
Historical notes will be presented in Section \ref{s22}.\\

Brownian motion $B(t)$, $t\ge 0$, is a stochastic process with many properties. 
It is at the same time Gaussian and  Markovian, has stationary increments and is self-similar. 
A process $X(t)$, $t\ge 0$, is said to be self-similar with self-similarity exponent $H$ if, 
for all $a\ge 0$, the processes $X(at)$, $t\ge 0$, and $a^{H}X(t)$, $t\ge 0$, 
have the same finite-dimensional distributions. Brownian motion is self-similar with exponent $H=1/2$. 
In contrast, fractional Brownian motion $B_H(t)$, $t\ge 0$, is Gaussian, has stationary increments, 
is self-similar with self-similarity exponent $0<H<1$, but is not Markovian, unless $H=1/2$, 
in which case the fractional Brownian motion becomes Brownian motion. When $1/2<H<1$, 
the increments of fractional Brownian motion have long-range dependence \cite{TaqquREV02}. \\ 

Because Brownian motion is Markovian with stationary increments, its finite-dimensional distributions 
can be obtained from the marginal density function
\begin{equation}
f_B(x,t)=\frac{1}{\sqrt{4\pi t}}e^{-x^2/4t},\;\; x\in \mathbb{R} 
\end{equation}
at time $t\ge 0$. This density function is the fundamental solution of the ``standard'' diffusion equation:
\begin{equation}
\partial_t u(x,t)=\partial_{xx}u(x,t),
\label{st0}
\end{equation} 
which in integral form reads:
\begin{equation}
u(x,t)=u_0(x)+\int_{0}^{t}\partial_{xx}u(x,s)ds,\;\; u_0(x)=u(x,0).
\label{st1}
\end{equation}
Thus, $f_B(x,t)$ is a solution of Eq. (\ref{st1}) with $u_0(x)=\delta(x)$, where $\delta(x)$ 
is the Dirac delta distribution. 
We allow, throughout the paper, functions to be distributions.
\begin{rem}
We follow the physics convention of not including the factor $1/2$ in Eq. (\ref{st0}). 
Therefore, in this paper, ``standard'' Brownian motion $B(t)$, $t\ge 0$, 
is such that, for each time $t\ge 0$, $B(t)\sim N(0,2t)$. The ``tilde'' notation $X\sim f_X(x)$ 
indicates that the random variable $X$ has the probability density function $f_X(x)$.  
\end{rem}
Our goal is to extend Eq. (\ref{st1}) to non-Markovian settings. 
We will consider non-local, fractional and stretched modifications of the diffusion equation. 
These modified equations will be called {\it Non-Markovian diffusion equations}, 
because, while they originate from a diffusion equation, the corresponding process, 
whose probability density function is a solution of these modified equations, will be typically non-Markovian.

To motivate the modifications, 
consider first the non-random process $l(t)=t$, $t\ge 0$, which depicts a non-random linear time evolution 
and let $f_l(\tau,t)$ denote its density function at time $t$. 
Therefore one has $f_l(\tau,t)=\delta(\tau-t)$ where $\delta(x)$ is the Dirac distribution. 
It is natural to interpret $f_l(\tau,t)$ as the fundamental solution of the standard forward drift equation:
\begin{equation}
\partial_tu(\tau,t)=-\partial_\tau u(\tau,t),\;\; \tau,t\ge 0,
\label{eqw}
\end{equation}
which in integral form reads:
\begin{equation}
u(\tau,t)=u_0(\tau)-\int_{0}^{t}\partial_\tau u(\tau,s)ds,\;\;\; u_0(\tau)=u(\tau,0).
\label{sfdr}
\end{equation}
The general solutions are of the form $u(\tau,t)=u_0(\tau-t)$ and thus, when $u_0(\tau)=\delta(\tau)$, 
the solution of Eq. (\ref{eqw}) is indeed $u(\tau,t)=\delta(\tau-t)$. 
Observe that the variable $\tau \ge 0$ plays the role of a space variable.\\ 

We will consider the following generalization of the forward drift equation (\ref{sfdr}) 
\begin{equation}
u(\tau,t)=u_0(\tau)-\int_{0}^{t}K(t-s)\partial_\tau u(\tau,s)ds,\;\; \tau,t\ge 0,
\label{stt4}
\end{equation}
where $K(t)$, with $t\ge 0$, is a suitable kernel chosen such that the fundamental solution of Eq. (\ref{stt4}) 
is a probability density function at each $t\ge 0$.
 We refer to Eq. (\ref{stt4}) as the {\it non-Markovian forward drift equation}.\\ 

The presence of the memory kernel $K$ in Eq. (\ref{stt4}) suggests a corresponding modification of the 
diffusion equation  (\ref{st1}). Namely, we will consider the equation:
\begin{equation}
u(x,t)=u_0(x)+\int_{0}^{t}K(t-s)\partial_{xx}u(x,s)ds,\;\; x\in \mathbb{R},\;\; t\ge 0.
\label{Mrkdi}
\end{equation}   
Its fundamental solution turns out to be:
\begin{equation}
f(x,t)=\int_{0}^{\infty}G(x,\tau)h(\tau,t)d\tau,
\label{fff}
\end{equation}
where 
\begin{equation}
G(x,t)=\frac{1}{\sqrt{4\pi t}}\exp(-x^2/4t),
\label{ggauss}
\end{equation}
and $h(\tau,t)$ is the fundamental solution of Eq. (\ref{stt4}).\\

The solution (\ref{fff}) is a marginal (one-point) probability density function. 
We will consider different random processes whose marginal probability density function coincides with it. 
As illustration, consider the following examples\footnote{In these examples we refer to facts which are justified later in the paper through forward references. The reader may want to focus at this
  point only on the examples and ignore the references.}. 
\begin{ese}\label{e1} 
If we choose:
\begin{equation} 
K(t)=\frac{t^{-1/2}}{\sqrt \pi},\;\;t\ge 0,
\label{kerf}
\end{equation}
then we have, see Eq. (\ref{pro3} and Eq. (\ref{curh}):
\begin{equation}
h(\tau,t)=\frac{1}{\sqrt{\pi t}}\exp\left(-\frac{\tau^2}{4t}\right),\;\; \tau \ge 0,\;\; t\ge 0,
\label{fu3}
\end{equation}
as the fundamental solution of Eq. (\ref{stt4}). Now consider the process
\begin{equation}
D(t)=B(l(t)),\;\; t\ge 0,
\label{stimp}
\end{equation}
where $B$ is a ``standard'' Brownian motion and $l(t)\ge 0$ is a random time-change (not necessarily increasing), 
independent of $B$, whose marginal density function is given by $h(\tau,t)$.
One possible choice for the random time process is simply:
$$
l(t)=|b(t)|,\;\; t\ge 0,
$$
where $b(t)$, $t \ge 0$, is a ``standard'' Brownian motion \cite{Feller1,hida}. 
Such a random time process $l(t)$, $t\ge 0$, 
is self-similar of order $H=1/2$.\footnote{
Another possible choice for a random time process with marginal density given by Eq. (\ref{fu3}) 
is the {\it local time} in zero of a ``standard'' Brownian motion \cite{beghin}. 
In this case the time-change process $l(t)$ is increasing.}
Let now $B(t)$, $t\ge 0$, be another ``standard'' Brownian motion independent of $b(t)$. 
Thus, the process (see also \cite{MerschertSub})  
\begin{equation}
D(t)=B(|b(t)|),\;\; t\ge 0,
\end{equation}
has marginal density defined by Eq. (\ref{fff}) with $h(\tau,t)$ 
given by Eq. (\ref{fu3}). 
\end{ese}
But, $D(t)$ is not the only process with density function $f(x,t)$, given by Eq. (\ref{fff}). 
For example, the process
\begin{equation}
Y(t)=\sqrt{|b(1)|}B_{1/4}(t),\;\; t\ge 0\,,
\label{fuf3}
\end{equation}
where $B_{1/4}$ is an independent fractional Brownian motion with self-similarity exponent $H=1/4$, 
has the same one-dimensional probability density functions as the previous process 
$D(t)$, $t\ge 0$, see Eq. (\ref{diste}) with $\beta =1/2$.  
\begin{ese}\label{ee2}
The fractional Brownian motion in Eq. (\ref{fuf3}) has a self-similarity exponent $H<1/2$. 
The increments of such a process are known to be negatively correlated 
\cite{MandelbrotJH76,MandelbrotNess68,TaqquREV02}. 
To allow for the presence of fractional Brownian motion $B_H(t)$ with $0<H<1$, 
we introduce a second (non-random) time-change $t\rightarrow g(t)$, 
where $g(0)=0$ and $g(t)$ is smooth and increasing, that is we consider the non-Markovian diffusion equation 
\begin{equation}
u(x,t)=u_0(x)+\int_{0}^{t}g'(s)K\left(g(t)-g(s)\right)\partial_{xx}u(x,s)ds.
\label{nmeq0}
\end{equation}
whose fundamental solution is now:
\begin{equation}
f(x,t)=\int_{0}^{\infty}G(x,\tau)h(\tau,g(t))d\tau,
\label{ff20}
\end{equation} 
where $h$ is the fundamental solution of Eq. (\ref{stt4}). 
If $K(t)$ is as in Eq. (\ref{kerf}) and $g(t)=t^{2\alpha}$, with $0<\alpha<2$, then the processes:
$$
\begin{array}{ll}
D(t)=B(|b(t^{2\alpha})|),\;\; t\ge 0,\\[0.3cm]
Y(t)=\sqrt{|b(1)|}B_{\alpha/2}(t),\;\; t\ge 0,
\end{array}
$$ 
have a marginal density function defined by Eq. (\ref{ff20}) with $h(\tau,t)$ as 
in Eq. (\ref{fu3}),
 which is the fundamental solution of Eq. (\ref{nmeq0}). 
 In this case $Y(t)$ is defined through an independent fractional Brownian motion $B_{\alpha/2}$ 
 with Hurst's parameter $H=\alpha/2$ and thus $0<H<1$. 
 This is a special case of Eq. (\ref{eqlb}).
\end{ese}

The preceding examples illustrate the themes pursued in the paper. 
We will focus, however, not only on power-like kernels such as those defined 
in Eq. (\ref{kerf}), but also on exponential-like kernels such as:
\begin{equation}
K(t)=e^{-at},\;\; a\ge 0.
\end{equation}
 
We also consider what happens when the Brownian motion $B(t)$, $t\ge 0$, 
is replaced by a more general linear (time-homogeneous) diffusion $Q(t)$, $t\ge 0,$ governed 
by the Fokker-Planck equation\footnote{Also known as the forward Kolmogorov equation.},
\begin{equation}
\partial_tu(x,t)=\mathcal{P}_xu(x,t),
\label{fpfp}
\end{equation}  
where $\mathcal{P}_x$ is a linear operator independent of $t$ acting on the variable $x\in \mathbb{R}$. In other words we consider the non-Markovian diffusion equation:
\begin{equation}
u(x,t)=u_0(x)+\int_{0}^{t}g'(s)K(g(t)-g(s))\mathcal{P}_xu(x,s)ds.
\label{m}
\end{equation} 
We show that its fundamental solution is:
\begin{equation}
f(x,t)=\int_{0}^{\infty}\mathcal{G}(x,\tau)h(\tau,g(t))d\tau,
\label{mfon}
\end{equation}
where $\mathcal{G}(x,t)$ is the fundamental solution of Eq. (\ref{m}), while $h(\tau,t)$ 
is the fundamental solution of Eq. (\ref{stt4}). 
We also provide explicit solutions when $\mathcal{P}_x$ is the differential operator associated 
with Brownian motion with drift, when it is associated with Geometric Brownian motion 
and when the kernel $K(t)$ is the power kernel and the exponential kernel.\\ 

In order not to dwell  on technicalities, 
we suppose implicitly, throughout the paper, that we have sufficient regularity conditions, 
to justify the algebraic manipulations that are performed. The paper is organized as follows:
\begin{itemize}

\item Historical notes are presented in Section \ref{s22}.
\item In Section \ref{St} we study the non-Markovian forward drift equation  (\ref{stt4}) 
and its corresponding random time process $l(t)$. 
We derive suitability conditions on the kernel $K(t)$. 
We end the section by noting that a self-similar time-change process, 
for instance with self-similarity parameter $H=\beta$, 
requires the choice $K(t)=Ct^{\beta-1}/\Gamma(\beta)$ with $0<\beta\le 1$. 
\item In Section \ref{s3} we study the non-Markovian diffusion equation (\ref{nmeq0}) 
and its solutions, and we discuss its various stochastic interpretations.
\item In Section \ref{thenot} we illustrate the fact that the stochastic representation is not unique.
\item In Section \ref{S3} we study the more general non-Markovian  Fokker-Planck equation 
and derive its solution Eq. (\ref{mfon}).
\item In Section \ref{S4} we go thorough several examples with
  $\mathcal{P}_xu(x,t)= \partial_{xx}u(x,t)$, that is, when the
  underlying diffusion proces is Brownian motion. 
  We consider non-Markovian diffusion equations, associated with the 
  $\beta${\it -power} kernel $K(t)=t^{\beta-1}/\Gamma(\beta)$, $0<\beta\le 1$, 
  and with the {\it exponential-decay} kernel $K(t)=e^{-at}$, $a\ge 0$. 
  We also consider different choices of the deterministic scaling function $g(t)$, 
  for example a logarithmic time scale $g(t)=\log(t+1)$ is considered.
\item In Section \ref{A} we focus on  applications when the  underlying diffusion process 
is not standard Brownian motion. 
We  consider the case of  Brownian motion with drift and  Geometric Brownian
  motion and we  study  the corresponding equations with  
  the {\it $\beta$-power} kernel and the {\it exponential-decay} kernel.
\item Section \ref{C} contains a summary and concluding remarks.
\end{itemize}    

\section{Historical notes}\label{s22}  
Non-Markovian equations like Eq. (\ref{Mrkdi}), or more generally
Eq. (\ref{m}), are often encountered when studying physical phenomena 
related to relaxation and diffusion problems in complex systems (see Srokowsky \cite{sro} for examples).\\

Equations of the type  (\ref{Mrkdi}) have been studied for example by  Kolsrud \cite{kolsrud}. 
He obtained Eq. \ref{fff}), but without providing specific examples. 
A similar study was done by Wyss \cite{wyss} who, however, focused only on power-like kernels $K(t)=Ct^{\beta-1}$.\\
 
Sokolov \cite{soko} (see also Srokowsky \cite{sro}), studied the non-Markovian equation
\begin{equation}
\partial_tP(x,t)=\int_0^t k(t-s)L_xP(x,s)ds,
\label{eqr}
\end{equation}
where $L_x$ is a linear operator acting on the variable $x$. 
He provided a formal solution in the form of Eq. \ref{mfon}). 
Observe, however, that our equation  (\ref{m}) differs from Eq. (\ref{eqr}), 
not only by the presence of the scaling function $g(t)$, but also by the choice of the memory kernel. 
Our kernel $K(t)$ and Sokolov's kernel $k(t)$ are related by the equation: 
\begin{equation}
K(t)=\int_{0}^tk(s)ds\Rightarrow \widetilde K(s)=\widetilde k(s)/s,\;\; s> 0,
\end{equation}
where the tilde indicates the Laplace transform, see Eq. (\ref{ltde}). 
The suitability conditions for these memory kernels are thus not the same 
(these conditions are developed in Section \ref{St}). 
For example, consider the simple {\it exponential-decay} kernel $e^{-at}$, $a\ge 0$. 
This choice of the kernel is ``safe'' in the context of Eq. (\ref{m}), i.e. for the choice $K(t)=e^{-at}$, 
but is ``dangerous'' if one considers Eq.  (\ref{eqr}) with the kernel $k(t)=e^{-at}$. 
In the case of Eq.  (\ref{m}), the {\it exponential-decay} kernel corresponds to a system for 
which non-local memory effects are initially negligible. 
In fact, $K(t)=e^{-at}\rightarrow 1$ as $t\rightarrow 0$ and thus the system appears Markovian at small times. 
On the other hand, the choice $k(t)=e^{-at}$ corresponds to the kernel $K(t)=a^{-1}(1-e^{-at})$ 
which for small times behaves like $t$. 
In this case Sokolov \cite{soko} noticed that the corresponding equations are only reasonable in 
a restricted domain of the model parameters and for certain initial and boundary conditions.\\ 

Our starting point is different from that of the previous authors. 
Instead of starting directly from the Fokker-Planck equations (\ref{fpfp}), 
we start from the forward drift equation (\ref{sfdr}) 
which is then generalized by introducing a memory kernel $K(t)$, Eq. (\ref{stt4}). 
One is then naturally led to the non-Markovian diffusion equations (\ref{nmeq0} 
and (\ref{m}) after the introduction of the scaling function $g(t)$. 
In fact, in specific cases, it is sometimes simpler to solve first the non-Markovian forward drift 
equation (\ref{stt4}) and then use the solution to solve the non-Markovian diffusion 
equation (\ref{nmeq0}) or (\ref{m}) by using (\ref{ff20}) or (\ref{mfon}). 
The form of the solution (\ref{ff20}) or  (\ref{mfon}) 
has now a ready-made interpretation. 
For example, in Eq. (\ref{mfon}) the function $\mathcal{G}(x,t)$ is the fundamental solution 
of the Markovian equation (\ref{fpfp})
 and the function $h(\tau,t)$ is the fundamental solution of the 
 non-Markovian equation (\ref{stt4}) and it is these two solutions that contribute 
 to Eq. (\ref{mfon}) which is the fundamental solution of the non-Markovian diffusion 
 equation (\ref{m}).\\ 

Furthermore, the form (\ref{ff20}) or  (\ref{mfon})  has a natural interpretation 
in terms of {\it subordinated} processes, see Eq. (\ref{stimp}). 
According to Whitmore and Lee \cite{lee}, the term ``subordination'' 
was introduced by Bochner \cite{Bochner55,Bochner62}. 
It  refers to processes of the form $Y(t)=X(l(t))$, $t\ge 0$, where $X(t)$, $t\ge 0$, 
is a Markov process and $l(t)$, $t\ge 0$, is a (non-negative) random time process independent of $X$. 
The marginal distribution of the subordinated process is clearly:
\begin{equation}
f_Y(x,t)=\int_0^\infty f_X(x,\tau)f_l(\tau,t)d\tau,\;\; t\ge 0,\;\; x\in \mathbb{R},
\label{subeq}
\end{equation}
where $f_X(x,t)$ and $f_l(\tau,t)$ represent the marginal density functions of the processes $X$ and $l$. 
Therefore, Eq. (\ref{ff20}) or Eq. (\ref{mfon}) can be interpreted in terms of subordinated processes,
 with Eq. (\ref{stt4}) characterizing the random time process $l(t)$ and
 Eq. (\ref{fpfp}) characterizing the Markov parent process $X(t)$.\\ 

The stochastic interpretation through subordinated processes, 
first suggested by Kolsrud, is very natural because $Y(t)=X(l(t))$  has a direct physical interpretation. 
For example, in equipment usage, $X(t)$ can be the state of a machine at time $t$ and $l(t)$ 
the effective usage up to time $t$. In an econometric study, $X(t)$ may be a model for the price of a stock 
at time $t$. 
If $l(t)$ measures the total economic activity up to time $t$, the price of the stock at time $t$ 
should not be described by $X(t)$ but by the subordinated process $Y(t)=X(l(t))$. 
The resulting subordinated process $Y(t)$ is in general non-Markovian. 
In this way, the non-local memory effects are attributable to the random time process 
$l(t)$ and to its dynamics  which is in general non-local in time, see Eq. (\ref{stt4}).\\

Note, however, that the solution of Eq. (\ref{m}) represents only the marginal (one-point) density function 
of the process and therefore cannot characterize the full stochastic structure of the process. 
As we note in the paper, there are also processes that are not subordinated processes 
that serve as stochastic models for non-Markovian diffusion equations like
Eq. (\ref{m}) or Eq. (\ref{eqr}). 

For example, consider in Eq. (\ref{Mrkdi}) the $\beta${\it -power} kernel 
$K(t)=t^{\beta-1}/\Gamma(\beta)$, with $0<\beta\le 1$. 
From a stochastic point of view, the fundamental solution of this equation, 
also called the time-fractional diffusion equation of order $\beta$, 
can be interpreted as the marginal density function of a self-similar stochastic processes with parameter $H=\beta/2$. 
This process, for example,  can be taken to be a subordinated process $Y(t)=B(l(t))$, with a suitable choice of the random time $l$. 
In Kolsrud \cite{kolsrud}, the random time $l$ is taken to be related to the local time 
of a $d=2(1-\beta)$-dimensional fractional Bessel process,
 while in Meerschaert et al. \cite{meerschaert} 
 (see also 
\cite{BarkaiChemPhys02,GorMaiHONNEF06,GorMaiVivCSF07,Kleinhans-Friedrich,saichev05,stani,GrigoliniRoccoWest99}), 
 in the context of a Continuous Time Random Walk (CTRW), 
 it is chosen to be the inverse of the totally skewed strictly $\beta$-stable process. 
 The interested reader is referred to the wide literature concerning the relationship 
between CTRW and non-Markovian diffusion equations and its applications. See for instance, 
\cite{BarkaiMetzlerKlafter, GorMaiINDIA02,HilferAntonPRE95, MetzlerKlafterPhysRep00,
MetzlerKlafterJPhysics04,ScalasGorenfloMainardiPRE04,WeissBOOK94,ZaslavskyPhysRep02,
ScalasGorenfloMainardiPhyA,MARAGO,GOMASCA} and references therein. 

Schneider \cite{Schneider1}, moreover, in a very general mathematical construction, 
introduced the so-called {\it Grey Brownian motion}. 
This process is a self-similar process with stationary increments which, as turns out, 
can be represented by $Y(t)=\Lambda_\beta B_H(t)$, $t\ge 0$, 
where $B_H$ is a fractional Brownian motion with $H=\beta/2$ and $\Lambda_\beta$ 
is a suitable chosen random variable independent of $B_H$ 
(see Mura et al. for details \cite{Mura1, Mura2}). 
This process has a marginal density function that evolves in time according to 
the time-fractional diffusion equation of order $\beta$. 
In this case the non-Markovian property is due to the presence of the fractional Brownian motion. 
As we show in the paper, long-range dependence can be made to appear through the time-scaling function 
$g(t)$, see Eq. (\ref{nmeq0}) and Example \ref{ee2}.
Figures 4, 5 and 6 display trajectories of the processes $D(t)$ and $Y(t)$ and corresponding density functions.

\section{The non-Markovian forward drift equation}\label{St}
We start with the following generalization of Eq. (\ref{sfdr}), namely: 
\begin{equation}
u(\tau,t)=u_0(\tau)-\int_{0}^{t}K(t-s)\partial_\tau u(\tau,s)ds,\;\; \tau,t\ge 0,
\label{fdr}
\end{equation}
where $K(t)$, with $t\ge 0$, is a suitable chosen kernel. 
We then choose a random time process $l(t)$ such that, for each $t\ge 0$, 
its marginal density $f_l(\tau,t)$ is the fundamental solution of Eq. (\ref{fdr}). 
Observe that Eq. (\ref{fdr}) is ``non-local'' because $u(\tau,t)$ 
involves $u(\tau,s)$ at all $0\le s\le t$. 
Equation (\ref{fdr}) will be called {\it non-Markovian forward drift equation}, 
see Section \ref{s1}, Eq. (\ref{stt4}).\\ 
  
It is convenient to work with Laplace transforms. We indicate by $\mathscr{L}\{\varphi (x,t);t,s\}$ the Laplace transform of the function $\varphi$ with respect to $t$ evaluated in $s \ge 0$, namely:
\begin{equation}
\mathscr{L}\{\varphi (x,t);t,s\}=\int_{0}^{\infty}e^{-ts}\varphi(x,t)dt,\;\; s\ge 0.
\label{ltde}
\end{equation}
If the function $\varphi$ depends only on the variable $t$ we write simply $\widetilde \varphi (s)$, because in this case there is no ambiguity concerning the integration variable. In particular we let $\widetilde K(s)$ denote the Laplace transform of the kernel $K$.
\begin{prop}\label{p1}
Let $f_l(\tau,t)$ denote the fundamental solution of Eq. (\ref{fdr}). 
Then,
\begin{equation}
\mathscr{L}\{f_l(\tau,t); t,s\}=\frac{1}{s\widetilde K(s)}\exp{\left(-\frac{\tau}{\widetilde K(s)}\right)},\;\; \tau,s\ge 0,
\label{lapt}
\end{equation}
and zero for $\tau <0$.
\end{prop}
\noindent
{\bf Proof}: we take the Laplace transform with respect to the variable $t$ in Eq. (\ref{fdr}):
\begin{equation}
\partial_\tau\widetilde u(\tau,s)=\frac{u_0(\tau)}{s\widetilde K(s)}-\frac{\widetilde u(\tau,s)}{\widetilde K(s)},
\label{peql}
\end{equation}
thus Eq. (\ref{lapt}) is a solution, in the distributional sense, when $u_0(\tau)=\delta(\tau)$. 
Indeed the general solution ofEq. \ref{peql}) with $u_0(\tau)=\delta(\tau)$ is:
$$
\varphi(\tau,s)=\frac{\theta(\tau)}{s\widetilde K(s)}\exp{\left(-\frac{\tau}{\widetilde K(s)}\right)}
+C\exp{\left(-\frac{\tau}{\widetilde K(s)}\right)},\;\;\; \tau \in \mathbb{R},
$$
where $C$ is a real constant and where 
\begin{equation}
\theta(x)=\left\{\begin{array}{ll} 1, & x\ge 0, \\ 0, & x<0 \end{array}\right.
\label{step}
\end{equation}
is the Heaviside's step function. Since we require $\varphi(\tau,t)=0$
for $\tau< 0$, we get $C=0$ i.e. Eq. (\ref{lapt}). $\Box$\\ 

\subsection{Suitability conditions on the kernel $K$}

We must choose the kernel $K$ such that the fundamental solution 
of Eq. (\ref{fdr}) is a probability density in $\tau \ge 0$. 
We observe that if $f_l(\tau,t)$ satisfies Eq. (\ref{fdr}) and Eq. (\ref{lapt}), 
then it is automatically normalized for each $t\ge 0$. 
In fact, for a function $\varphi(x,t)$ for which it is always possible to change the order of integration, 
one has:
\begin{equation}
\int_{\mathbb{R}} \varphi(x,t)dx=1 \Longleftrightarrow \int_{\mathbb{R}} \widetilde \varphi(x,s)dx=s^{-1}.
\label{lapnorm}
\end{equation}
Since Eq. (\ref{lapt}) satisfies the right-hand side of
Eq. (\ref{lapnorm}), we get $\int_{\mathbb{R}_+}f_l(\tau,t)d\tau=1$. 
One still needs, however, to choose the kernel $K$ such that  $f_l(\tau,t)\ge 0$ for all $\tau,t \ge 0$. \\ 

In order to get a suitable condition on the kernel $K$, we make use of the notion of completely monotone function. 
Recall that a function $\varphi(t)$ is {\it completely monotone} 
if it is non-negative and possesses derivatives of any order and:
\begin{equation}
(-1)^k\frac{d^k}{dt^k}\varphi(t)\ge 0,\;\;\; t> 0,\;\; k\in \mathbb{Z}_+=\{0,1,2,\dots\}.
\end{equation}
We observe that as $t \rightarrow 0$, the limit of $d^{k} \varphi(t)/dt^k$ may be finite or infinite.
Typical non-trivial examples are $\varphi(t)=\exp(-at)$, with $a> 0$, $\psi(t)=1/t$ and $\phi(t)=1/(1+t)$. 
It is easy to show that if $\varphi$ and $\psi$ are completely monotone then their product $\varphi\psi$ is as well. 
Moreover, if $\varphi$ is completely monotone and $\psi$ 
is positive with first derivative completely monotone then the function $\varphi(\psi)$ is completely monotone.\\ 

We have the following characterization of completely monotone functions \cite{Feller2}:
\begin{lemma}
A function $\varphi(s)$, defined on the positive real line, is completely monotone 
if and only if is of the form:
$$
\varphi(s)=\int_{0}^{\infty}e^{-ts}F(dt),\;\; s\ge 0,
$$  
where $F$ is a finite or infinite non-negative measure on the positive real semi-axis. 
\end{lemma}
Hence, to ensure that $f_l(\tau,t)\ge 0$ for all $\tau,t\ge 0$, 
it is enough to require  that the function defined in Eq. (\ref{lapt}) must be  completely monotone, 
as a function of $s$, for any $\tau \ge 0$, 
and thus that the kernel $K$ satisfies the following:
\subsubsection*{Suitability conditions}
\begin{enumerate}
\item $s\widetilde K(s)$ is positive with first derivative completely monotone, 
\item $1/\widetilde K(s)$ is positive with first derivative completely monotone.
\end{enumerate}
Indeed, we can view Eq. \ref{lapt}) as the product of the two completely monotone functions 
$1/u$  and $\exp(-\tau u)$, 
the first evaluated at $u=s\widetilde K(s)$ and the second evaluated at $u=1/\widetilde{K}(s)$.  

\subsection{Examples}

\begin{ese}[{\it $\beta$-power kernel}]\label{es1}
If we choose:
$$
K(t)=\frac{t^{\beta-1}}{\Gamma(\beta)},
$$
we get $\widetilde K(s)=s^{-\beta}$. 
In this case $s\widetilde K(s)=s^{1-\beta}$ 
is positive and has first derivative $(1-\beta)s^{-\beta}$ completely monotone if and only if $0<\beta\le 1$. 
Moreover, $1/\widetilde K(s)=s^{\beta}$ 
is positive with first derivative $\beta s^{\beta-1}$ completely monotone if and only if 
$0<\beta\le 1$. Therefore, a good choice for the kernel $K$ is:
\begin{equation}
K(t)=\frac{t^{\beta-1}}{\Gamma(\beta)},\;\;\; 0<\beta\le 1.
\end{equation}
\end{ese}
\begin{ese}[{\it Exponential-decay kernel}]\label{esexp}
Choosing:
\begin{equation}
K(t)=\exp(-at),\;\; a\ge 0,
\end{equation}
we get $s\widetilde K(s)=s/(s+a)$ 
which is positive with first derivative $a(s+a)^{-2}$ completely monotone for any $a\ge 0$. 
Moreover, $1/\widetilde K(s)=(s+a)$ is positive if  $a \ge 0$ with first derivative completely monotone.
\end{ese}
\begin{ese}[{\it $\beta$-power with exponential-decay kernel}]
Choosing: 
\begin{equation}
K(t)=\frac{t^{\beta-1}}{\Gamma(\beta)}\exp(-a t),\;\; 0<\beta\le 1,\;\; a \ge 0,
\end{equation}
we have $\widetilde K(s)=(s+a)^{-\beta}$. 
Therefore, $s\widetilde K(s)=s(s+a)^{-\beta}$ 
which is positive if $a \ge 0$ with first derivative $(s+a)^{-\beta}(1-\beta s(s+a)^{-1})$ 
completely monotone if $0<\beta \le 1$. 
Moreover, $1/\widetilde K(s)=(s+a)^{\beta}$ is positive if $a \ge 0$ 
with first derivative $\beta(s+a)^{\beta-1}$ completely monotone if $0<\beta\le 1$. 
\end{ese}
The following theorem states that a self-similar random time process $l(t)$, $t\ge 0$, 
is associated with the  kernel $K(t)$ in Example \ref{es1}:
\begin{teor}\label{ppp}
If the time-change process $l(t)$, $t \ge 0$, 
is self-similar (for instance of order $H=\beta$), with marginal probability density $f_l(\tau,t)$ 
satisfying Eq. (\ref{lapt}), then we must have:
\begin{equation}
K(t)=C\frac{t^{\beta-1}}{\Gamma(\beta)},\;\; 0<\beta\le 1,
\label{ssK}
\end{equation}
for some positive constant $C$. 
\end{teor}
\noindent
{\bf Proof}: The self-similarity condition entails that for any $\tau,t \ge 0$ and for any $a >0$:
$$
a^{-\beta}f_l(a^{-\beta}\tau,t)=f_l(\tau,at).
$$
If we take the Laplace transform and set $\widetilde f(\tau,s)=\mathcal{L}\{f_{l}(\tau,t);t,s\}$, we have:
$$
a^{-\beta}\widetilde f_l(a^{-\beta}\tau,s)=\frac{1}{a}\widetilde f_l\left (\tau,\frac{s}{a}\right ).
$$
Using Eq. \ref{lapt}) we get that for any $\tau,s\ge 0$ and $a >0$: 
$$
\frac{a^{-\beta}}{\widetilde K(s)}\exp \left(-\frac{a^{-\beta}\tau}{\widetilde K(s)}\right)=\frac{1}{\widetilde K(\frac{s}{a})}\exp \left(-\frac{\tau}{\widetilde K(\frac{s}{a})}\right).
$$
Since this relation is valid for any choice of $\tau\ge 0$ and $s\ge 0$, 
putting $\tau=0$ and $s=a$, we get:
$$
\frac{a^{-\beta}}{\widetilde K(a)}=\frac{1}{\widetilde K(1)}.
$$
Thus, for any $a>0$:
$$
\widetilde K(a)=\widetilde K(1) a^{-\beta},
$$
which is the Laplace transform ofEq. \ref{ssK}). 
If we add moreover the condition of complete monotonicity we find: $0<\beta\le 1$ 
as indicated in Example \ref{es1}. $\Box$ 

\section{Non-Markovian diffusion equation}\label{s3}
We focus here on the non-Markovian diffusion equation (\ref{nmeq0}) introduced in the first section. 
There are two ingredients:
\begin{enumerate}
\item The fundamental solution ofEq. \ref{fdr}), denoted here by $h(\tau,t)$ 
and defined by Eq. (\ref{lapt}).
\item The fundamental solution $G(x,t)$, defined by
Eq. (\ref{ggauss}), of the standard diffusion equation which 
is the one-dimensional density of the ``standard'' Brownian motion. 
\end{enumerate}
The following theorem combines these 
two ingredients and provides the fundamental solution of a corresponding non-Markovian diffusion equation.
\begin{teor}\label{t1}
Let $h(\tau,t)$ denote the fundamental solution of Eq. \ref{fdr}),
 so that by Proposition \ref{p1}, one has:
\begin{equation}
\mathscr{L}\{h(\tau,t); t,s\}=
\frac{1}{s\widetilde K(s)}\exp{\left(-\frac{\tau}{\widetilde K(s)}\right)},\;\; \tau,s\ge 0,
\label{laptg}
\end{equation}
for a suitable choice of $K$. Let $g$ be a strictly increasing function with $g(0)=0$ and 
let $G(x,t)$ be defined by Eq. (\ref{ggauss}). Then, 
\begin{equation}
f(x,t)=\int_{0}^{\infty}G(x,\tau)h(\tau,g(t))d\tau,
\label{ff2}
\end{equation} 
 is the fundamental solution of the non-Markovian diffusion equation:
\begin{equation}
u(x,t)=u_0(t)+\int_{0}^{t}g'(s)K\left(g(t)-g(s)\right)\partial_{xx}u(x,s)ds.
\label{nmeq}
\end{equation}
\end{teor}
\noindent
{\bf Proof}: see Section \ref{S3}.  $\Box$\\ \\
We have immediately the following:
\begin{cor}\label{cor1}
If $H(x,t)$ is a solution of the standard diffusion equation with initial condition $H(x,0)=u_0(x)$, then the function:
\begin{equation}
u(x,t)=\int_{0}^{\infty}H(x,\tau)h(\tau,g(t))d\tau
\label{gensol}
\end{equation}
is a solution of Eq. (\ref{nmeq}). 
\end{cor}
\noindent
{\bf Proof}: If, for any $t\ge 0$, the function $f(x,t)$ 
defined in Eq. (\ref{ff2}) is the fundamental solution of Eq. (\ref{nmeq}) then a general solution is given by:
$$
u(x,t)=\int_{\mathbb{R}}f(x-y,t)u_0(y)dy=
\int_{\mathbb{R}}\int_{0}^{\infty}G(x-y,\tau)u_0(y)h(\tau,g(t))\,d\tau \,dy
$$
$$
=\int_{0}^{\infty}\left(
\int_{\mathbb{R}}G(x-y,\tau)u_0(y)dy\right)h(\tau,g(t))d\tau=
\int_{0}^{\infty}H(x,\tau)h(\tau,g(t))d\tau.\;\;\; \Box 
$$
We observe that:
\begin{enumerate}
\item The equation (\ref{laptg}) states that $h(\tau,t)$ 
is the fundamental solution of Eq. (\ref{fdr}). 
\item While $G(x,t)$ is the fundamental solution of the standard diffusion equation 
obtained when $u_0(x)=\delta(x)$, the general solution, denoted $H(x,t)$ in the above theorem, 
results from a general initial condition $u_0(x)$. 
\end{enumerate}
Many physical phenomena, especially related to relaxation processes in complex systems, 
are described by non-Markovian ``master equations'' like Eq. (\ref{nmeq}). $K(t)$ is a memory kernel 
and $g(t)$ is just a ``time-scaling`` function. 
Such equations are often argued by phenomenological considerations and 
 can be more or less rigorously derived starting from a microscopic description \cite{denz,zwan, kenkre, sro}. 

\section{The stochastic representation is not unique}\label{thenot}
The solution of the non-Markovian diffusion equation can be viewed as the marginal density function 
of the subordinated process, see Eq. (\ref{stimp})
$$
D(t)=B(l(g(t))),\;\; t \ge 0,
$$
since its marginal density is:
$$
f_D(x,t)=\int_{0}^{\infty}G(x,\tau)f_l(\tau,g(t))d\tau.
$$
Here, for each $t\ge 0$, $D(t)\sim f_D(x,t)$, $B(t)\sim G(x,t)$ and $l(t)\sim f_l(\tau,t)$. 
In the notation of Theorem \ref{t1}, we have $f_D(x,t)=f(x,t)$ and $f_l(\tau,t)=h(\tau,t)$. 
The Laplace transform of $f_l(\tau,t)$ with respect to $t$ is given by Eq. (\ref{laptg}).\\
 
This stochastic representation is not unique (see Example \ref{e1}, Example \ref{ee2} 
and examples below). 
Indeed, the non-Markovian diffusion equation characterizes only the marginal, 
that is one-point, probability density function. 
However, processes with a different dependence structure can have the same marginal density $f(x,t)$. 
Additional requirements could be imposed so as to specify the stochastic model more precisely. 
\begin{ese}\label{s2}
If we require the random time process $l_\beta(t)$, $t\ge 0$, to be self-similar of order $\beta$, 
then in view of Theorem \ref{ppp}, the kernel must be chosen as in Eq. (\ref{ssK}) and 
we must have $0<\beta\le 1$. 
We will study this case more in details in Section \ref{S4}. 
Here we just observe that if we consider a ``standard'' fractional Brownian motion 
$B_{\beta/2}$ of order $\beta/2$, then $f(x,t)$ is also the marginal distribution of 
\begin{equation}
Y(t)=\sqrt{l_{\beta}(1)}B_{\beta/2}(t),
\end{equation} 
where $B_{\beta/2}(t)$ is assumed to be independent of $l_\beta(1)$.\\ \\
In fact, because $l_\beta(t)$, $t\ge 0$, is self-similar of order $H=\beta$, one has:
\begin{equation}
D(t)=B(l_{\beta}(t))=^{\!\!\!\!{}^d}\sqrt{l_\beta(t)}B(1)=^{\!\!\!\!{}^d}\sqrt{ l_\beta(1)} 
t^{\beta/2}B(1)=^{\!\!\!\!{}^d}\sqrt{l_\beta(1)} t^{\beta/2}B_{\beta/2}(1)=^{\!\!\!\!{}^d}
\sqrt{l_\beta(1)} B_{\beta/2}(t)=Y(t),
\label{diste}	
\end{equation}
where $=^{\!\!\!\!{}^d}$\, denotes here the equality of the marginal distributions.
\end{ese}
Both $D(t)$, $t\ge 0$, and $Y(t)$, $t\ge 0$, are self-similar processes with 
Hurst's exponent $H=\beta/2$. 
However, while $Y(t)$, $t\ge 0$, 
has always stationary increments, this is not in general true in the case of the process $D(t)$, $t\ge 0$.

\section{Non-Markovian Fokker-Planck equation}\label{S3}
We considered up until now processes of the type $B(l(g(t)))$, where $B$ is a 
``standard'' Brownian motion. 
What happens if we replace $B$ by a more general diffusion? 
Namely, what happens if instead of starting with the standard diffusion
 equation (\ref{st0}) we start with a more general Markovian Fokker-Planck equation:
\begin{equation}
\partial_t u(x,t)=\mathcal{P}_x u(x,t),\;\; x\in \mathbb{R},\;\; t\ge 0,
\label{fPl}
\end{equation}
where $\mathcal{P}_x$ is a linear operator, independent of $t$, acting on the variable $x$? 
We have the following generalization of Theorem \ref{t1}:
\begin{teor}\label{t2}
Suppose that $h(\tau,t)$ is a probability density function satisfyingEq. \ref{lapt})
\begin{equation}
\mathscr{L}\{h(\tau,t); t,s\}=\frac{1}{s\widetilde K(s)}
\exp{\left(-\frac{\tau}{\widetilde K(s)}\right)},\;\; \tau,s\ge 0,
\label{lapt3}
\end{equation}
for a suitable choice of $K$. Let $g$ be a strictly increasing function with $g(0)=0$ and $\mathcal{G}(x,t)$ 
be the fundamental solution of Eq. (\ref{fPl}). Then
the fundamental solution of the integral equation:
\begin{equation}
u(x,t)=u_0(t)+\int_{0}^{t}g'(s)K\left(g(t)-g(s)\right)\mathcal{P}_xu(x,s)ds
\label{nmeq2}
\end{equation}
is 
\begin{equation}
f(x,t)=\int_{0}^{\infty}\mathcal{G}(x,\tau)h(\tau,g(t))d\tau.
\label{ff3}
\end{equation} 
\end{teor}
\noindent
We provide two versions of the proof. The first starts with the solution $f(x,t)$ in
Eq. (\ref{ff3}) and verifies that it satisfies Eq. (\ref{nmeq2}). 
The second starts from the partial integro-differential equation (\ref{nmeq2}) 
and derives the solution $f(x,t)$ under certain assumptions stated below Eq. (\ref{LFD}). \\ \\
{\bf Proof 1}: For first we observe that
\begin{equation}
\mathscr L\{f(x,t);g(t),s\}=\frac{1}{s\widetilde K(s)}\mathscr{L}\{\mathcal{G}(x,t);t,\widetilde K(s)^{-1}\}.
\label{lappp2}
\end{equation}
With the change of variables $g(s)=z$, we write:
\begin{equation}
u(x,g^{-1}(w))=u_0(x)+\int_{0}^{w}K\left(w-z\right)\mathcal{P}_xu(x,g^{-1}(z))dz,\;\; w=g(t).
\label{rew2}
\end{equation}
We want to show that Eq. (\ref{ff3}) with the choice (\ref{lapt3}) 
solves Eq. (\ref{nmeq2}). 
If we take the Laplace transform of Eq. (\ref{nmeq2}) using Eq. (\ref{rew2}), we get:
$$
\mathscr{L}\{u(x,t);g(t),s\}=\frac{u_0(x)}{s}+\widetilde K(s)\mathcal{P}_x\mathscr{L}\{u(x,t);g(t),s\}
$$
that is:
\begin{equation}
s\mathscr{L}\{u(x,t);g(t),s\}-u_0(x)=s\widetilde K(s)\mathcal{P}_x\mathscr{L}\{u(x,t);g(t),s\}.
\label{onlap2}
\end{equation}
Now, if we substitute on Eq. (\ref{onlap2}) a solution of the form  (\ref{ff3}),
\begin{equation}
u(x,t)=\int_{0}^{\infty}\mathcal{H}(x,\tau)h(\tau,g(t))d\tau,
\label{fff3}
\end{equation}
we have:
$$
\widetilde K(s)^{-1}\mathscr{L}\{\mathcal{H}(x,t);t,\widetilde K(s)^{-1}\}=
u_0(x)+\mathcal{P}_x\mathscr{L}\{\mathcal{H}(x,t);t,\widetilde K(s)^{-1}\}
$$
i.e. we have, with obvious notations:
$$
\tau \widetilde {\mathcal{H}}(x,\tau)=u_0(x)+\mathcal{P}_x\widetilde {\mathcal{H}}(x,\tau),
$$
in which one readily recognizes the Laplace transform of the Markovian Fokker-Planck equation 
with the same initial condition $u_0(x)$. Therefore:
$$
\partial_t \mathcal{H}(x,t)=\mathcal{P}_x \mathcal{H}(x,t),\;\;\; \mathcal{H}(x,0)=u_0(x).
$$
This argument shows not only that Eq. (\ref{ff3}) is the fundamental solution 
of Eq. (\ref{nmeq2}), but also that a general solution is given by
Eq. (\ref{fff3}) (see Corollary \ref{cor1}). 
This result is summarized in Corollary \ref{cor2} (see below).\\ \\
{\bf Proof 2}: We now start from Eq.(\ref{nmeq2}) 
and we use integral transforms in order to get the fundamental solution. 
Let $\mathcal{F}$ denote the Fourier transform operator and let:
$$
(\mathcal{F} \varphi)(k,t)=\widehat \varphi(k,t)=\int_{\mathbb{R}}e^{ikx}\varphi(x,t)\,dx\,.
$$  
Since $\widehat u_0(k)=1$, and since 
$(\mathcal{F}\mathcal{P}_xu)(k,t)=(\mathcal{F}\mathcal{P}_x\mathcal{F}^{-1}\mathcal{F}u)(k,t)=
\widehat {\mathcal{P}}_k\widehat u(k,t)$, 
where $\widehat{\mathcal{P}}_k=(\mathcal{F}\mathcal{P}_x\mathcal{F}^{-1})_k$ 
denotes the Fourier transform of the operator $\mathcal{P}_x$, we have:
$$
\widehat u(k,g^{-1}(w))=1+\int_{0}^{w}K(w-z)\widehat {\mathcal{P}}_k\widehat u(k,g^{-1}(z))dz.
$$
Taking the Laplace transform we have:
$$
\mathscr{L}\{{\widehat {u}}(k,g^{-1}(w));w,s\}=s^{-1}+\widehat {\mathcal{P}}_k\widetilde K(s)\mathscr{L}\{{\widehat{u}}(k,g^{-1}(w));w,s\},
$$
which is the same as:
$$
\mathscr{L}\{{\widehat {u}}(k,t);g(t),s\}=s^{-1}+\widehat {\mathcal{P}}_k\widetilde K(s)\mathscr{L}\{{\widehat{u}}(k,t);g(t),s\}.
$$
Therefore:
$$
\left(\widetilde K(s)^{-1}-\widehat {\mathcal{P}}_k\right)\mathscr{L}\{{\widehat {u}}(k,t);g(t),s\}=s^{-1}\widetilde K(s)^{-1}.
$$
Denoting $1(k)=1$, we have: 
\begin{equation}
\mathscr{L}\{{\widehat {u}}(k,t);g(t),s\}=\frac{1}{s\widetilde K(s)}\left(\widetilde{K}(s)^{-1}-\widehat {\mathcal{P}}_k\right)^{-1}1(k),
\label{LFD}
\end{equation}
where we suppose that the operator $\left(\widetilde{K}(s)^{-1}-\widehat {\mathcal{P}}_k\right)^{-1}$ is well defined and acts on the constant function $1(k)=1$.\\ \\
Observe that the Fokker-Planck equation (\ref{fPl}) is obtained from Eq. (\ref{nmeq2}) 
by setting $K(t)=1$, for each $t\ge 0$, that is $\widetilde K(s)=s^{-1}$, and $g(t)=t$, 
for each $t\ge 0$. In this case Eq. (\ref{LFD}) becomes: 
\begin{equation}
\mathscr{L}\{{\widehat {\mathcal{G}}}(k,t);t,s\}=(s-\widehat {\mathcal{P}}_k)^{-1}1(k).
\end{equation}
where $\mathcal{G}(x,t)$ is the fundamental solution. Taking the inverse Fourier transform, we get:
\begin{equation}
\mathscr{L}\{\mathcal{G}(x,t);t,s\}=\mathscr{F}^{-1}\left\{(s-\widehat {\mathcal{P}}_k)^{-1}1(k)\;;\;k,x\right\},
\label{ffo}
\end{equation}
where:
\begin{equation}
\mathscr{F}^{-1}\left\{\varphi(k,s)\;;\;k,x\right\}=\frac{1}{2\pi}\int_{\mathbb{R}}e^{-ikx}\varphi(k,s)dk.
\end{equation}
Replacing $s$ by $\widetilde K(s)^{-1}$ in Eq. (\ref{ffo}), one has:
\begin{equation}
\mathscr{L}\{\mathcal{G}(x,t);t,\widetilde K(s)^{-1}\}=
\mathscr{F}^{-1}\left\{(\widetilde K(s)^{-1}-\widehat {\mathcal{P}}_k)^{-1}1(k)\;;\;k,x\right\}.
\label{stst}
\end{equation} 
Going back to Eq. (\ref{LFD}) and inverting the Fourier transform we obtain in view of
Eq. (\ref{stst}):
$$
\mathscr{L}\{u(x,t);g(t),s\}=\frac{1}{s\widetilde K(s)}\mathscr{F}^{-1}
\left\{\left(\widetilde{K}(s)^{-1}-\widehat {\mathcal{P}}_k\right)^{-1}1(k)\;;\;
k,x\right\}=\frac{1}{s\widetilde K(s)}\mathscr{L}\{\mathcal{G}(x,t);t,\widetilde{K}(s)^{-1}\}.
$$
that is Eq. (\ref{lappp2}).$\;\;\Box$
\begin{rem} 
If the Markovian process is a Brownian motion one has 
$\mathcal{P}_x=\displaystyle\frac{\partial^2}{\partial x^2}$. 
The Fourier transform 	of $\mathcal{P}_x$ is $\widehat{\mathcal{P}}_k=-k^2$ and Eq. (\ref{LFD}) becomes:
$$
\mathscr{L}\{{\widehat {u}}(k,t);g(t),s\}=\frac{1}{s\widetilde K(s)}
\left(\widetilde{K}(s)^{-1}+k^2\right)^{-1}1(k),
$$  
where 
$$
\left(\widetilde{K}(s)^{-1}+k^2\right)^{-1}1(k)=\frac{1}{\left(\widetilde{K}(s)^{-1}+k^2\right)},
$$
which is well defined because $\widetilde K(s)^{-1}$ is positive.
\end{rem}
\begin{cor}\label{cor2}
If $\mathcal{H}(x,t)$ is a general solution of the Markovian Fokker-Planck equation (\ref{fPl}) with initial condition $\mathcal{H}(x,0)=u_0(x)$, then the function:
\begin{equation}
u(x,t)=\int_{0}^{\infty}\mathcal{H}(x,\tau)h(\tau,g(t))d\tau
\label{gensol2}
\end{equation}
is a general solution of Eq. (\ref{nmeq2}). 
\end{cor}	
>From a stochastic point of view, $f(x,t)$ could be seen as the marginal distribution at time $t$ 
of the subordinated process:
\begin{equation}
\mathcal{D}(t)=Q(l(g(t)))
\label{sQ}
\end{equation}
where $Q$ is the diffusion governed by the Fokker-Planck equation (\ref{fPl}) and $l(t)$ is the 
random time process, independent of $Q(t)$, with marginal distributions defined by $h(\tau,t)$.

\section{Examples involving standard Brownian motion}\label{S4}
In the following examples, we consider stochastic models where  
the operator $\mathcal{P}_x$ in Eq. (\ref{fPl}) is $\partial_{xx}$, 
namely the operator corresponding to standard Brownian motion. 
We  will study more general operators in the next section. 
We shall choose various kernels $K(t)$ and various stretching functions $g(t)$. 
We let $h(\tau,t)$ denote the fundamental solution of the
 non-Markovian forward drift equation  (\ref{fdr}).
  Since the corresponding stochastic models are not unique, 
  we will mainly focus on the subordinated process $B(l(t))$, $t\ge 0$. 
  However, we also give examples of other  appropriate stochastic models.
\subsection{Time-fractional diffusion equation}\label{sstfde}
Let $g(t)=t$. Consider the $\beta$-{\it power} kernel:
\begin{equation}
K(t)=\displaystyle\frac{t^{\beta-1}}{\Gamma(\beta)},\;\;0<\beta \le 1, 
\label{cho1}
\end{equation}
and let $h(\tau,t)$ denote the fundamental solution of the non-Markovian forward drift 
equation (\ref{fdr}) with kernel (\ref{cho1}).
\begin{rem}
In view of Theorem \ref{ppp},
such a kernel arises if one requires $h(\tau,t)$ to be the marginal
density function of a self-similar random time process $l(t)$ of order $\beta$.
\end{rem}
\noindent
InsertingEq. \ref{cho1}) in Eq. (\ref{nmeq}) we obtain the following equation:
\begin{equation}
u(x,t)=u_0(t)+\frac{1}{\Gamma(\beta)}\int_{0}^{t} \left(t-s\right)^{\beta-1}\partial_{xx}u(x,s)\,ds\,,
\label{TFDE1}
\end{equation}
which is sometimes called the {\it time-fractional diffusion equation} \cite{schneider3,LUMAPA}. 
In view of Theorem \ref{t1}, the fundamental solution is:
$$
f(x,t)=\int_{0}^{\infty}G(x,\tau)h(\tau,t)\,d\tau,
$$
where $h(\tau,t)$ satisfies:
\begin{equation}
\mathscr{L}\{h(\tau,t); t,s\}=s^{\beta-1}e^{-\tau s^{\beta}},\;\; \tau,s\ge 0.
\label{fraclap}
\end{equation}
Such a function $h(\tau,t)$ can be expressed as:
\begin{equation}
h(\tau,t)=t^{-\beta}M_{\beta}(\tau t^{-\beta}),
\label{loct}
\end{equation}
where $M_{\beta}(r)$, is defined for $0<\beta< 1$ by the power series \cite{MainardiCal1, MainardiCal2}:
\begin{equation}
\begin{array}{ll}
M_{\beta}(r) &= \,{\displaystyle \sum_{k=0}^{\infty}\frac{(-r)^k}{k!\Gamma\left[-\beta k+(1-\beta	)\right]}} \\
&=\, {\displaystyle \frac{1}{\pi}}\, 
{\displaystyle \sum_{k=0}^{\infty}\frac{(-r)^k}{k!}\Gamma\left[(\beta(k+1))\right]
\sin\left[\pi\beta(k+1)\right]},\;\;\; r\ge 0.
\end{array}
\label{functionM}  
\end{equation}
The above series defines a transcendental function (entire of order $1/(1 - \beta)$) \cite{GorMai}.
\begin{figure}[!h]
\begin{center} 
\includegraphics[keepaspectratio=true,height=9cm]{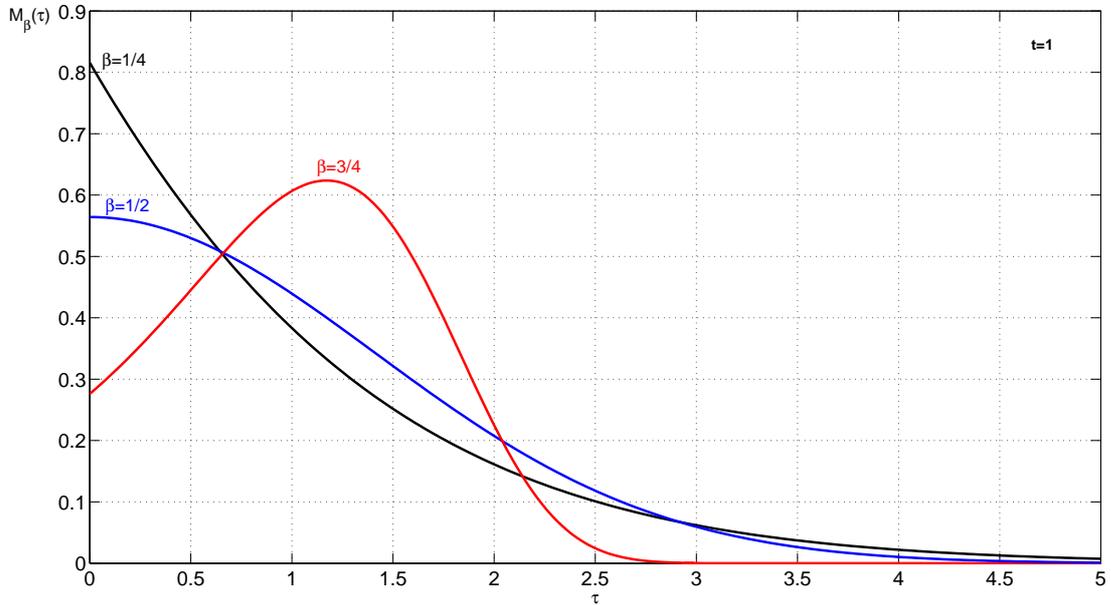}
\end{center}
\caption{Plot of the density function $h(\tau,t)=t^{-\beta}M(\tau t^{-\beta})$ at time $t=1$, 
for different values of the parameter $\beta=[1/4,1/2,3/4]$. \label{hplot}} 
\end{figure}
\begin{rem}\label{o3} 
The function $h(\tau,t)$ in Eq. (\ref{loct}) represents the fundamental solution of 
the time-fractional forward drift equation (see also \cite{GorMaiHONNEF06}):
\begin{equation}
u(\tau,t)=u_0(\tau)-\frac{1}{\Gamma(\beta)}\int_{0}^{t}(t-s)^{\beta-1}\partial_\tau u(\tau,s)ds.
\end{equation} 
This equation reduces to the standard drift equation when $\beta\rightarrow 1$.
\end{rem}
\subsubsection{Properties of the $M$-function}
\begin{figure}[!h]
\begin{center} 
\includegraphics[keepaspectratio=true,height=9cm]{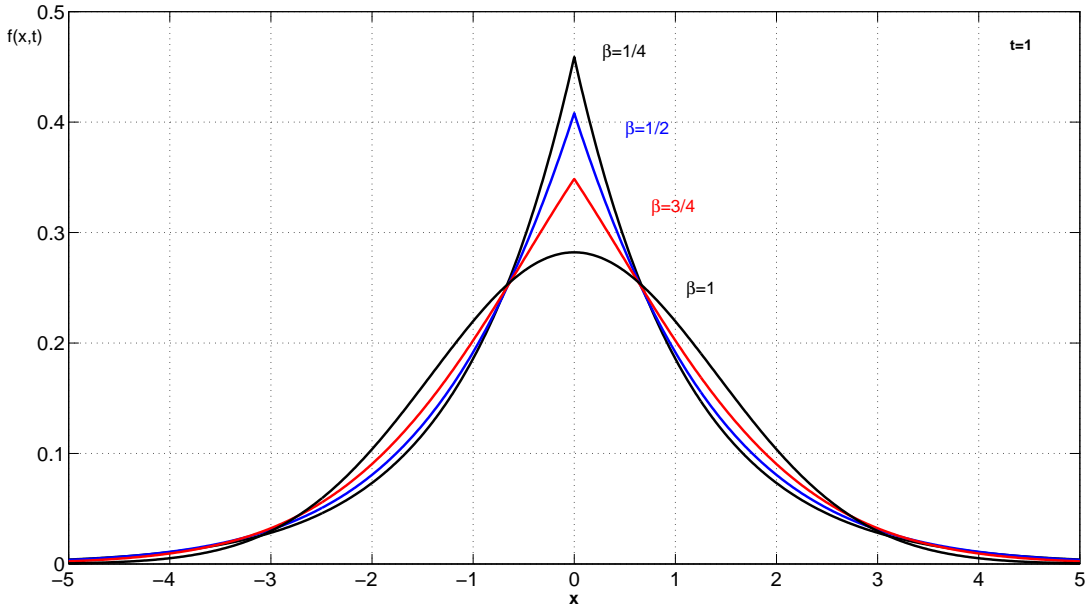}
\end{center}
\caption{Plot of the density function $f(x,t)$ given by Eq. (\ref{gbm}) at time $t=1$, 
for different values of the parameter $\beta=[1/4,1/2,3/4,1]$. 
For $\beta=1$ one recovers the standard Gaussian density  (\ref{st3}). \label{fplot}} 
\end{figure}

\begin{figure}[!h]
\begin{center} 
\includegraphics[keepaspectratio=true,height=9cm]{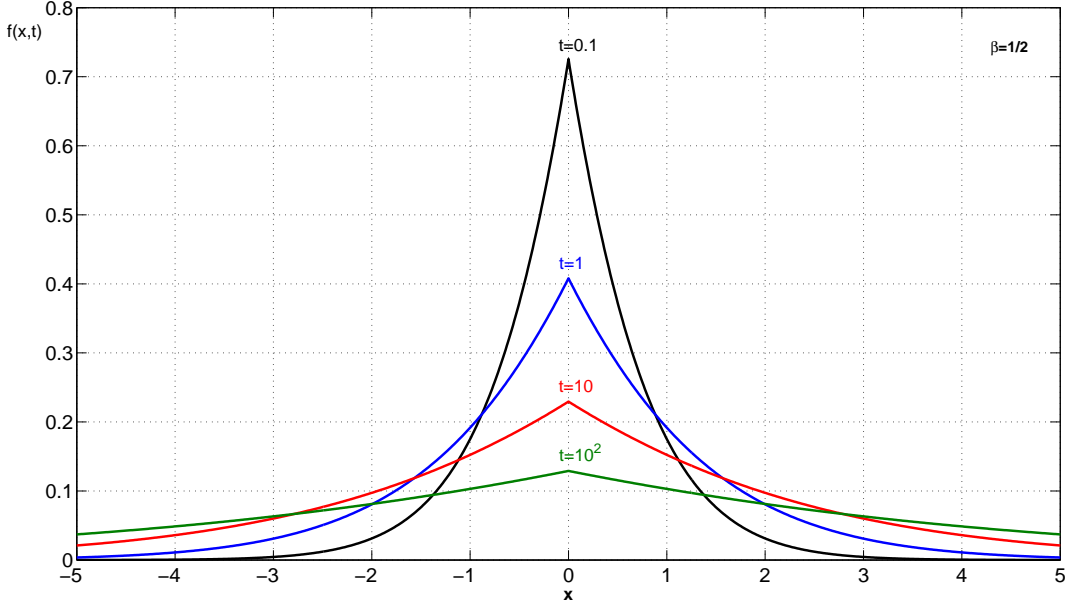}
\end{center}
\caption{Plot of the density function $f(x,t)$ for fixed $\beta=1/2$, at different times $t=[0.1,1,10,10^2]$. 
\label{fplott}} 
\end{figure}

It is useful to recall some important properties of the $M$-function \cite{MaPaGo03, GorMai}. 
These are best expressed in terms of the function 
\begin{equation}
\mathcal{M}_\beta(\tau,t)=t^{-\beta}M_\beta(\tau t^{-\beta}),
\end{equation}
defined for any $\tau,t \ge 0$ and $0<\beta <1$.
\begin{enumerate}
\item  The Laplace transform of $\mathcal{M}_\beta(\tau,t)$ with respect to $t$ is:
\begin{equation}
\mathscr{L}\{\mathcal{M}_\beta(\tau,t);t,s\}=s^{\beta-1}e^{-\tau s^{\beta}},\;\; \tau,s\ge 0.
\label{pro1}
\end{equation} 
\item The above equation suggests that in the singular limit $\beta \rightarrow 1$ one has:
\begin{equation}
\mathcal{M}_1(\tau,t)=\delta(\tau-t),\;\; \tau,t\ge 0.
\label{pro2}
\end{equation}
\item If $\beta=1/2$:
\begin{equation}
\mathcal{M}_{1/2}(\tau,t)=\frac{1}{\sqrt{\pi t}}\exp(-\tau^2/4t),\;\; \tau,t \ge 0.
\label{pro3}
\end{equation}
\item The $M$-function is a particular case of a Fox $H$-function \cite{mainardiH,schneider3}. We indicate with
\begin{equation}
\mathscr{M}\{\varphi(x);x,u\}=\int_{0}^{\infty}\varphi(x)x^{u-1}dx,
\end{equation}
the Mellin transform of a function $\varphi(x)$, $x\ge 0$, with respect to $x$ evaluated in $u\ge 0$. The Fox $H$-function 
$$
H^{m,n}_{p,q}(z)=H^{m,n}_{p,q}\left(z\Big{|}\begin{array}{l}(a_i,\alpha_i)_{i=1,\dots, p}\\ (b_j,\beta_j)_{j=1,\dots, q}\end{array}\right),
$$
is characterized by its Mellin transform as follows:
\begin{equation}
\mathscr{M}\{H^{\;m,n}_{p,q}(z);z,u\}=\frac{A(u)B(u)}{C(u)D(u)},
\label{HMell}
\end{equation}
with
$$
\begin{array}{ll}
A(u)=\displaystyle\prod_{i=1}^{m}\Gamma(b_j+\beta_j u), & B(u)=\displaystyle\prod_{j=1}^{n}\Gamma(1-a_j-\alpha_j u),\\ C(u)=\displaystyle\prod_{i=m+1}^{q}\Gamma(1-b_j-\beta_j u), & D(u)=\displaystyle\prod_{j=n+1}^{p}\Gamma(a_j+\alpha_j u).
\end{array}
$$
Here: $1\le m\le q$, $0\le n\le p$, $\alpha_j,\beta_j > 0$ and $a_j, b_j\in \mathbb{C}$ 
(see \cite{Fox,Mathai-SaxenaH-BOOK78,SrivastavaH-BOOK82} for more details).\\

Starting from Eq. (\ref{pro1}) 
and skipping to the Mellin transform, 
it is easy to show that we have the following relation:
\begin{equation}
\mathcal{M}_\beta(\tau,t)=t^{-\beta}H^{1,0}_{1,1}\left(\tau t^{-\beta}\Big{|}\begin{array}{l}(1-\beta,\beta)
\\ (0,1)\end{array}\right),\;\; \tau,t\ge 0,\;\; 0<\beta<1.
\label{Mfox}
\end{equation}
\item Using the representation Eq. (\ref{Mfox}) and Eq. (\ref{HMell}) 
we have for any $\eta,\beta\in (0,1)$, see also \cite{MaPaGo03}: 
\begin{equation}
\mathcal{M}_\nu(x,t)=
\int_{0}^{\infty}\mathcal{M}_{\eta}(x,\tau)\mathcal{M}_{\beta}(\tau,t)d\tau,\;\; \nu=\eta\beta\,\;\; x\ge 0.
\label{pro4}
\end{equation} 
\end{enumerate}
The expression  (\ref{loct}) for the function $h(\tau,t)$ follows from Eq. (\ref{pro1}), that is:
\begin{equation}
h(\tau,t)=\mathcal{M}_\beta(\tau,t),\;\;\tau,\,t \ge 0.
\label{curh}
\end{equation}
Moreover, when $\beta \rightarrow 1$,Eq. \ref{pro2}) 
gives $h(\tau,t)=\delta(\tau-t)$ as expected (see Remark \ref{o3}). 
Comparing Eq. (\ref{ggauss}) and Eq. (\ref{pro3}) one observes that:
\begin{equation}
G(x,t)=\frac{1}{2}\mathcal{M}_{1/2}(|x|,t).
\label{st3}
\end{equation}
Using Theorem \ref{t1} and Eq. (\ref{pro4}) together with 
Eq. (\ref{curh}) and  Eq. (\ref{st3}) 
we recover the fundamental solution of the time-fractional diffusion equation \cite{LUMAPA}:
$$
f(x,t)=
\int_{0}^{\infty}G(x,\tau)h(\tau,t)d\,\tau
=\frac{1}{2}\int_{0}^{\infty}\mathcal{M}_{1/2}(|x|,\tau)\mathcal{M}_\beta(\tau,t)\,d\tau 
$$
\begin{equation}
=\frac{1}{2}\mathcal{M}_{\beta/2}(|x|,t)=\frac{1}{2}t^{-\beta/2}M_{\beta/2}(|x|t^{-\beta/2}).
\label{gbm}
\end{equation}
Several plots of the $M$-function are presented: in Figure \ref{hplot} 
the function $h(\tau,t)=\mathcal{M}_\beta(\tau,t)$ is drawn at a fixed time $t=1$ 
and for different values of the parameter $\beta$; 
in Figure \ref{fplot} is presented the plot of $f(x,t)=\frac{1}{2}\mathcal{M}_{\beta/2}(|x|,t)$ 
at a fixed time $t=1$ and for different values of 
$\beta$; in Figure \ref{fplott} is shown the time evolution of $f(x,t)$ for fixed $\beta=1/2$. 
\subsubsection{Stochastic interpretations of the solution}
From a stochastic point of view, the function $h(\tau,t)$ in Eq. (\ref{loct}) can 
be regarded as the marginal distribution of 
$$
l_{\beta}(t),\;\; t\ge 0,
$$
where $l_{\beta}(t)$, $t\ge 0$, is  an $H$-ss random time with $H=\beta$. 
We have that for each integer $m\ge 0$:
\begin{equation}
E(l_\beta(t)^m)=\frac{m!}{\Gamma(\beta m+1)}t^{\beta m}.
\label{lmom} 
\end{equation}
In fact, from Eq. (\ref{fraclap}), for each integer $m\ge 0$, we have :
$$
\int_0^{\infty} \tau^m s^{\beta-1}e^{-\tau s^\beta}d\tau=m!s^{-m\beta-1},
$$
which, inverting the Laplace transform, gives Eq. (\ref{lmom}).\\

For instance, with the suitable conventions \cite{donati}, 
$l_{\beta}(t)$, $t\ge 0,$ can be viewed as the local time in zero at time $t$ 
of a $d=2(1-\beta)$-dimensional Bessel process \cite{molchanov}. 
The function $f(x,t)$ in Eq. (\ref{gbm}) is then the marginal density function of
$$
D(t)=B\left(l_{\beta}(t)\right),
$$
which is self-similar with $H=\beta/2$. 
In this case, because $l_{\beta}(t)$ is self-similar of 
order $\beta$, we immediately have an example of a different process with 
the same marginal distribution of $D(t)$ (see Example \ref{s2}). 
In fact, if we consider a ``standard'' fractional Brownian motion $B_{\beta/2}$ of order $\beta/2$, 
then $f(x,t)$ can also be seen as the marginal distribution of 
\begin{equation}
Y(t)=\sqrt{l_{\beta}(1)}B_{\beta/2}(t),
\end{equation} 
where $B_{\beta/2}(t)$ is assumed to be independent of $l_\beta(1)$ 
(see Example \ref{s2}). 
The process $Y(t)$, $t\ge 0$, is called {\it grey Brownian motion} \cite{Schneider1}.\\ 
\begin{figure}[!t]
\begin{center} 
\includegraphics[keepaspectratio=true,height=10cm]{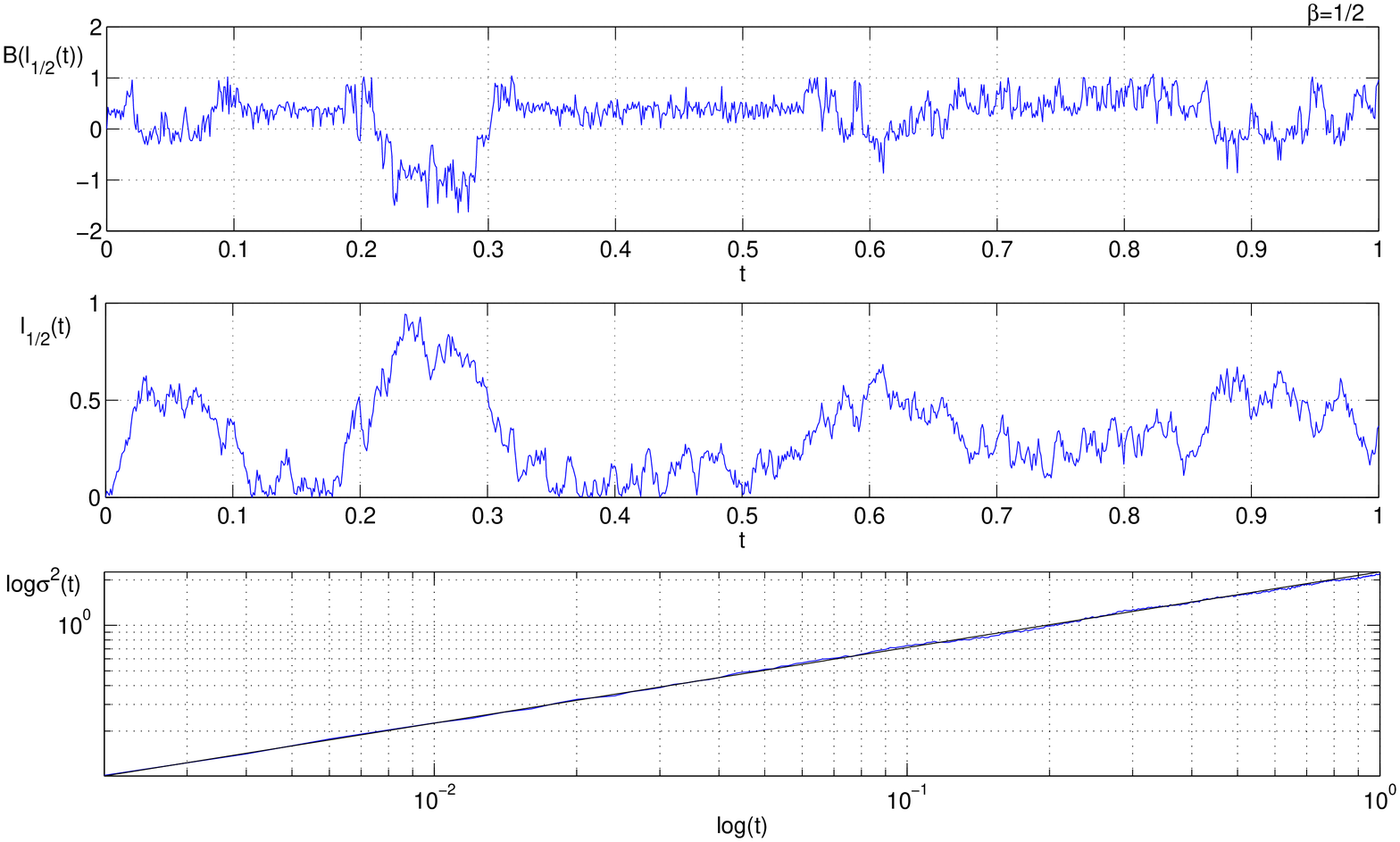}
\end{center}
\caption{Trajectory of the process $B(l_\beta(t))$ (top panel), with $0<t<1$ and $\beta=1/2$. 
The random time process is chosen to be $l_{1/2}(t)=|b(t)|$ where $b(t)$ is 
a ``standard'' Brownian motion (see Example \ref{e1}). 
The corresponding trajectory of the random time process is presented in the middle panel. 
The estimated variance, computed on a sample of dimension $N=5000$, 
is presented in logarithmic scale in the bottom panel and fits perfectly the theoretical curve $2t^{1/2}/\Gamma(3/2)$. \label{fracb}}
\end{figure}

From Eq. (\ref{lmom}) one can derive immediately all the moments for the processes $D(t)$ and $Y(t)$. 
For any integer $m\ge 0$
\begin{equation}
\left\{
\begin{array}{ll}
E(D(t)^{2m+1})=E(Y(t)^{2m+1})=0;\\[0.3cm]
E(D(t)^{2m})=E(Y(t)^{2m})=\displaystyle\frac{2m!}{\Gamma(\beta m+1)}t^{\beta m}.
\end{array}
\right.
\end{equation}
Because $0<\beta <1$, the variance grows slower than linearly with respect to time. 
In this case one speaks about {\it slow anomalous diffusion}.  
Moreover, the increments of the fractional Brownian motion $B_{\beta/2}(t)$ 
do not have long-range dependence. 
In contrast, the next example allows for the presence of long-range dependence through the 
introduction of a scaling function $g(t)=t^{\alpha/\beta}$ (see also Example \ref{ee2}).
\begin{figure}[!t]
\begin{center} 
\includegraphics[keepaspectratio=true,height=9cm]{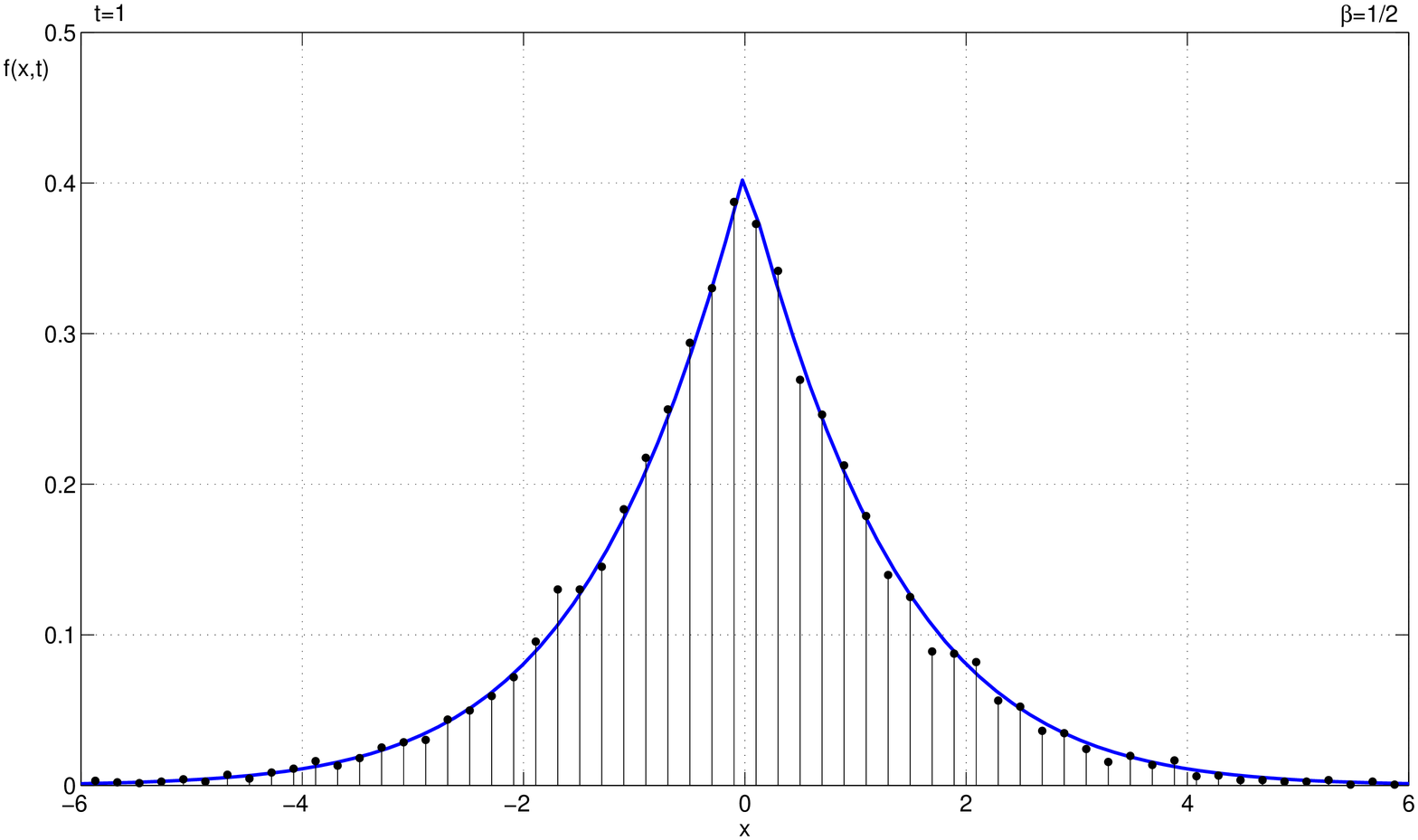}
\end{center}
\caption{Marginal density function $f(x,t)=\frac{1}{2}\mathcal{M}_{1/4}(|x|,t)$ 
of the process $B(l_{1/2}(t))$ at time $t=1$ and $x\in [-5,5]$. 
The histogram is evaluated over $N=10^4$ simulated trajectories of the process $B(|b(t)|)$ (Figure \ref{fracb}).}
\end{figure}
\subsection{``Stretched'' time-fractional diffusion equation}\label{sttfra}
If in the setup of Section \ref{sstfde}, where the kernel $K(t)$ is given by
Eq. (\ref{cho1}), we introduce a scaling time 
$$
g(t)=t^{\alpha/\beta}
$$ 
with $\alpha>0$, then the integral equation (\ref{TFDE1}) is replaced by Eq. (\ref{nmeq}), namely
\begin{equation}
u(x,t)=u_0(t)+\frac{1}{\Gamma(\beta)}\frac{\alpha}{\beta}\int_{0}^{t} s^{\frac{\alpha}{\beta}-1}\left(t^{\frac{\alpha}{\beta}}-s^{\frac{\alpha}{\beta}}\right)^{\beta-1}\partial_{xx}u(x,s)ds.
\label{ggbmeq}
\end{equation}
Therefore, using Eq. \ref{loct}):
$$
h(\tau,g(t))=g(t)^{-\beta}M_{\beta}(\tau g(t)^{-\beta})=t^{-\alpha}M_{\beta}(\tau t^{-\alpha}),
$$
and, using Eq. (\ref{gbm}), the fundamental solution $\overline f(x,t)$ of
Eq. (\ref{ggbmeq})  reads:
\begin{equation}
\overline f(x,t)=f(x,g(t))=\frac{1}{2}t^{-\alpha/2}M_{\beta/2}(|x|t^{-\alpha/2}),\;\; t\ge 0.
\label{ggbm}
\end{equation}
The function $\overline f(x,t)$, $t\ge 0$, is the marginal distribution of the process
$$
D(t)=B\left(l_{\beta}(t^{\alpha/\beta}))\right),\;\; t\ge 0.
$$
The time-change process  $l_{\beta}(t^{\alpha/\beta})$ is self-similar of order $H=\alpha$ and the process $D(t)$ is then self-similar with $H=\alpha/2$. In the case $0<\alpha<2$, the function $\overline f(x,t)$ is also the marginal density of
\begin{equation}
\mathcal{Y}(t)=\sqrt{l_{\beta}(1)}B_{\alpha/2}(t),\;\; t\ge 0,\;\; 0<\alpha<2,
\label{eqlb}
\end{equation}
where $B_{\alpha/2}(t)$ is a ``standard'' fBm of order $H=\alpha/2$ independent of $l_{\beta}(1)$. The process $\mathcal{Y}(t)$, $t\ge 0$, is called {\it generalized grey Brownian motion} \cite{Mura1}.\\ 

In this case, for any integer $m\ge 0$:
\begin{equation}
\left\{
\begin{array}{ll}
E(D(t)^{2m+1})=E(\mathcal{Y}(t)^{2m+1})=0;\\[0.3cm]
E(D(t)^{2m})=E(\mathcal{Y}(t)^{2m})=\displaystyle\frac{2m!}{\Gamma(\beta m+1)}t^{\alpha m}.
\end{array}
\right.
\end{equation}
We have slow diffusion when $0<\alpha<1$ 
(the variance grows slower than linearly in time) and fast diffusion when $1<\alpha<2$ 
(the variance grows faster than linearly in time). 
In this case the increments of the process $\mathcal{Y}(t)$ exhibit long-range dependence.
\begin{figure}[!t]
\begin{center} 
\includegraphics[keepaspectratio=true,height=9cm]{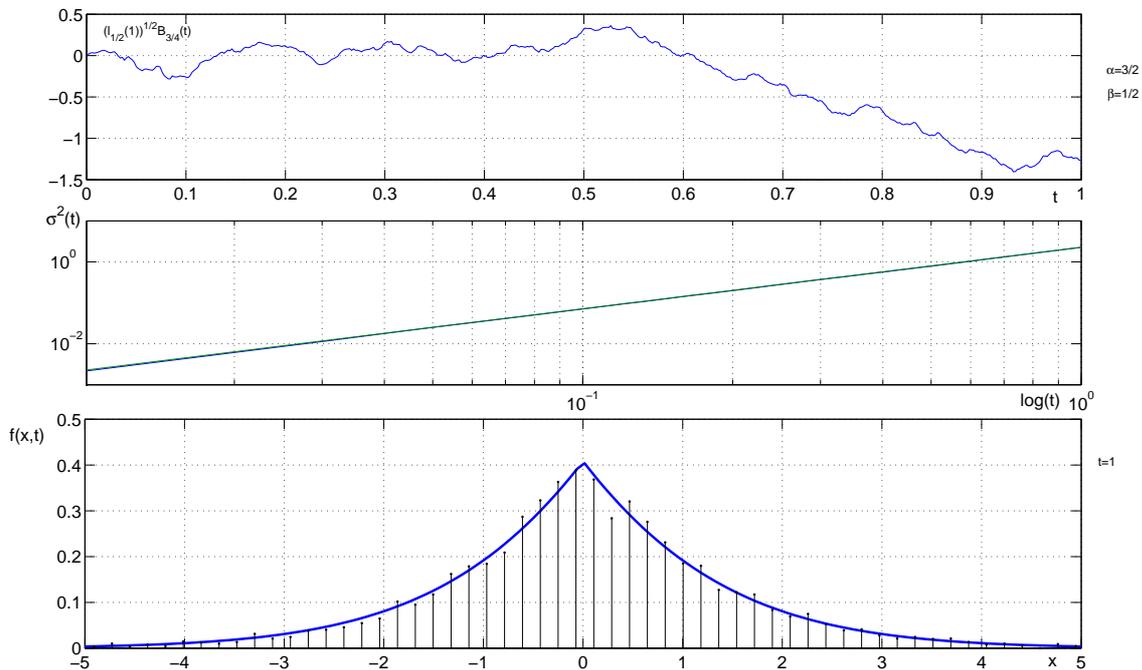}
\end{center}
\caption{Trajectory of the process $\sqrt{l_{\beta}(1)}B_{\alpha/2}(t)$ 
(top panel), with $0<t<1$, $\beta=1/2$ and $\alpha=3/2$. 
The random variable $l_{1/2}(1)$ is Gaussian, see Eq. (\ref{pro3}). 
The estimated variance, computed on a sample of dimension $N=5000$, 
is presented in logarithmic scale in the middle panel together with the theoretical curve $2t^{3/2}/\Gamma(3/2)$. 
In the bottom panel the histogram, evaluated over a sample of $N=10^4$ 
trajectories, fits the exact marginal density Eq. (\ref{ggbm}) at time $t=1$.}
\end{figure}
\subsection{Exponential-decay kernel}
\begin{figure}[!t]
\begin{center} 
\includegraphics[keepaspectratio=true,height=9cm]{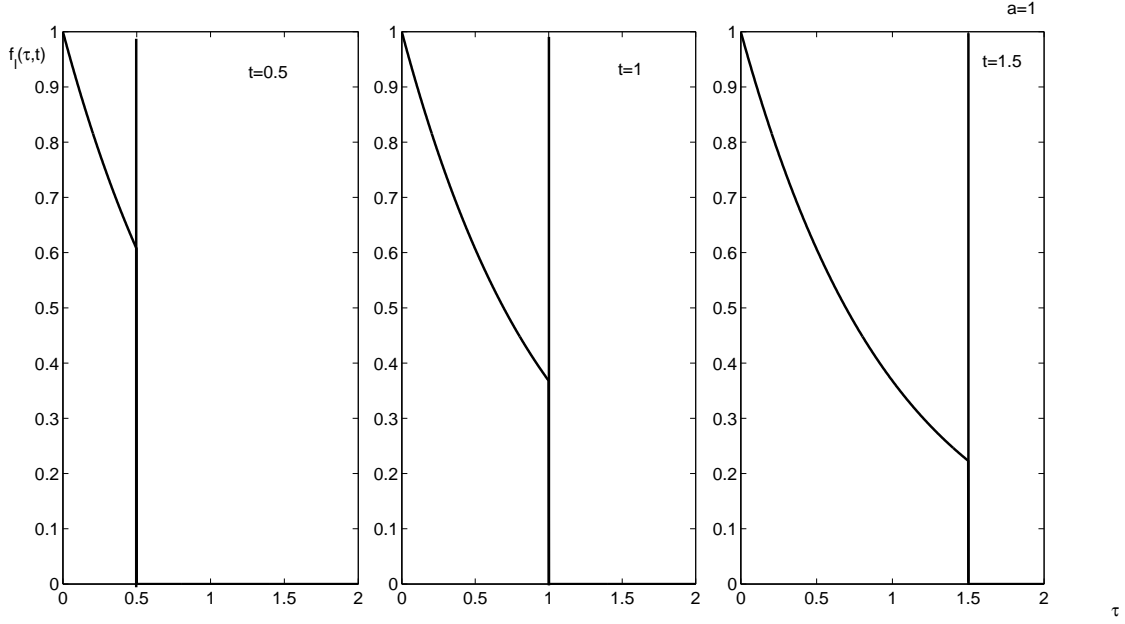}
\end{center}
\caption{Plots of the marginal density of the random time $f_l(\tau,t)$Eq. \ref{exptime}) 
as a function of $\tau$ at times $t=[0.5,1,1.5]$, and with $a=1$. 
The vertical line corresponds to a point mass (delta function). \label{flexp}} 
\end{figure}

Let $g(t)=t$.  With the exponential-decay kernel:
\begin{equation}
K(t)=\exp(-at),\;\; a\ge 0,\;\; t\ge 0,
\end{equation}
we obtain the following equation:
\begin{equation}
u(x,t)=u_0(x)+\int_{0}^{t}e^{-a(t-s)}\partial_{xx}u(x,s)ds.
\label{expdiff}
\end{equation}
In this case $\widetilde K(s)=(s+a)^{-1}$ 
and the marginal distribution of the random time process $l(t)$, $t\ge 0$, is defined by Eq. (\ref{lapt}):
$$
\mathscr{L}\{f_l(\tau,t);t,s\}=\frac{s+a}{s}e^{-\tau(s+a)},\;\; \tau\ge 0.
$$
Therefore,
\begin{equation}
f_l(\tau,t)=e^{-\tau a}\left (\delta(\tau-t)+a\theta (t-\tau)\right)=e^{-t a}\delta(\tau-t)+ae^{-\tau a}\theta(t-\tau),
\label{exptime}
\end{equation}
where $\theta(x)$ is the step function (\ref{step}). 
A graphical representation of the time evolution of $f_l(\tau,t)$ is presented in Figure \ref{flexp}. 
\begin{rem}
The function $f_{l}(\tau,t)$ defined in Eq. \ref{exptime}) is the fundamental solution, 
in the sense of distributions, of the {\it ``exponential'' forward drift} equation:
$$
u(\tau,t)=u_0(\tau)-\int_{0}^{t}e^{-a(t-s)}\partial_\tau u(\tau,s)ds.
$$
This follows from Proposition \ref{p1}. 
To check it directly we note that $f_{l}(\tau,0)=\delta(\tau)$ and for any $t >0$:
$$
-\int_{0}^{t}e^{-a(t-s)}\partial_\tau f_{l}(\tau,s)ds=
-\int_{0}^{t}e^{-a(t-s)}\partial_\tau\left(e^{-as}\delta(\tau-s)+ae^{-a\tau}\theta(s-\tau)\right)ds
$$
$$
=-\int_{0}^{t}e^{-a(t-s)}\left(-e^{-as}\delta'(\tau-s)+
ae^{-a\tau}\delta(s-\tau)-a^2e^{-a\tau}\theta(s-\tau)\right)\,ds
$$
$$
=e^{-at}\delta(\tau-t)+ae^{-at}\theta(t-\tau)+ae^{-a(t+\tau)}\theta(t-\tau)(e^{at}-e^{a\tau})
$$
$$
=e^{-at}\delta(\tau-t)+ae^{-a\tau}\theta(t-\tau)=f_{l}(\tau,t),
$$
where we have used the fact that: 
$$
\int_{0}^{t}\delta'(\tau-s)ds=\delta(t-\tau).
$$
We observe that when $a\rightarrow 0$ we recover the forward drift equation (\ref{sfdr}) 
and indeed $f_{l}(\tau,t)=\delta(\tau-t)$.
\end{rem}
\noindent
As noted in Example \ref{esexp}, Eq. (\ref{exptime}) actually defines a probability density for any $t\ge 0$.
The following proposition provides its moments. 
\begin{prop}
For each integer $m\ge 0$ one has:
\begin{equation}
E(l(t)^m)=\frac{m!}{a^m}\left(1-e^{-at}\right)+e^{-at}\left(t^m-\sum_{k=1}^{m}\frac{m!}{k!}t^{k}a^{k-m}\right).
\label{explmom}
\end{equation}
\end{prop}
\noindent
{\bf Proof}:  for any $t\ge 0$, we must evaluate:
$$
\int_{0}^{\infty}\tau^mf_l(\tau,t)d\tau=e^{-at}t^{m}+a\int_{0}^{t}\tau^me^{-a\tau}d\tau,
$$
where we have used Eq. (\ref{exptime}). 
In order to evaluate the exponential integral in the above equation we write:
$$
a\int_{0}^{t}\tau^me^{-a\tau}d\tau=(-1)^m a\partial^m_a\left[(1-e^{-at})(a^{-1})\right]=
(-1)^ma\sum_{k=0}^{m}\left(\!\!\begin{array}{cc} m\\ k\end{array}\!\!\right)\partial^k_a(1-e^{-at})
\partial^{m-k}_a(a^{-1})
$$
$$
=\sum_{k=0}^m(-1)^k\frac{m!}{k!}a^{k-m}\partial^k_a(1-e^{-at})=
\frac{m!}{a^m}(1-e^{-at})-\sum_{k=1}^{m}\frac{m!}{k!}t^ka^{k-m}e^{-at}
$$
thus one has Eq. \ref{explmom}). $\Box$\\

The function $f_l(\tau,t)$ can be written:
\begin{equation}
f_l(\tau,t)=e^{-at}\delta(\tau-t)+(1-e^{-at})\varphi(\tau,t),\;\; \tau,t\ge 0,\;\; a\ge 0,
\end{equation}
where:
\begin{equation}
\varphi(\tau,t)=a\frac{e^{-a\tau}\theta(t-\tau)}{1-e^{-at}},\;\; \tau,t\ge 0,\;\; a\ge 0.
\label{exptime2}
\end{equation}
Because $f_l(\tau,t)$ is a probability density, then so is $\varphi(\tau,t)$. 
The corresponding random time process $l(t)$, $t\ge 0$, 
can then be chosen to be:
\begin{equation}
l(t)=b_t t+(1-b_t)j(t),\;\; t\ge 0,
\label{randexp}
\end{equation}  
where $b_t$, $t\ge 0$, is a stochastic process such that, for any
fixed $t\ge 0$, $b_t$ is a Bernoulli random variable with
$Pr(b_t=1)=e^{-at}$ and $Pr(b_t=0)=1-e^{-at}$, and
$j(t)$, $t\ge 0$, is a stochastic process, independent of $b_t$, 
with marginal distribution given by $\varphi(\tau,t)$. 
\begin{rem} 
The random time $l(t)$ defined by Eq. (\ref{randexp}) cannot be increasing everywhere. 
This is due to the fact that $b_t$ and $j(t)$ are independent and $Pr(j(t)< t)=1$ for any $t\ge 0$. 
Indeed, suppose that $l(t)$ is increasing. This implies that for any $t\ge 0$ and $\epsilon >0$: 
$$
1=Pr(l(t+\epsilon)\ge l(t)\big{|}\,b_t=1)= Pr(l(t+\epsilon)\ge t)
$$ 
$$
=Pr(l(t+\epsilon)\ge t\big{|}\,b_{t+\epsilon}=1)Pr(b_{t+\epsilon}=1)+Pr(l(t+\epsilon)\ge t\big{|}\,b_{t+\epsilon}=0)Pr(b_{t+\epsilon}=0)
$$
$$
=e^{-a(t+\epsilon)}+\left(1-e^{-a(t+\epsilon)}\right)Pr(j(t+\epsilon)\ge t)
$$
$$
=1-\left(1-e^{-a(t+\epsilon)}\right)Pr(j(t+\epsilon)< t)
$$
therefore taking $\epsilon \rightarrow 0$ we get $1=e^{-at}$ with $a,t\ge 0$, 
which is a contradiction as soon as $a\neq 0$ and $t >0$.
\end{rem}
On the other hand, a trivial example of an increasing process with marginal distribution 
given by Eq. (\ref{exptime}) is:
\begin{equation}
\overline l(t)=\min(X,t),\;\; t\ge 0,
\label{overtime}
\end{equation}
where $X$ is an exponentially distributed random variable: $X\sim ae^{-a \tau}$, $\tau\ge 0$.\\ 

We now turn to Eq. \ref{expdiff}). We have the following result:
\begin{prop}
The fundamental solution of Eq. (\ref{expdiff}) is:
\begin{equation}
f(x,t)=e^{-at}G(x,t)+(1-e^{-at})\phi(x,t),
\label{brownber}
\end{equation}
with:
\begin{equation}
\phi(x,t)=\frac{\sqrt{a}}{4(1-e^{-at})}\left\{e^{x\sqrt a}\textrm{\rm Erf}\left(\frac{x}
{2\sqrt t}+\sqrt{at}\right)-e^{-x\sqrt a}\textrm{\rm Erf}\left(\frac{x}{2\sqrt t}-\sqrt{at}\right) 
 -2\sinh (|x|\sqrt a)\right\},
\end{equation}
where {\rm Erf}$(x)=\frac{2}{\sqrt{\pi}}\int_{0}^{x}e^{-y^2}dy$ 
and where {\rm Erf$(-x)=-$Erf$(x)$}.
\end{prop}
\begin{figure}[!h]
\begin{center} 
\includegraphics[keepaspectratio=true,height=9cm]{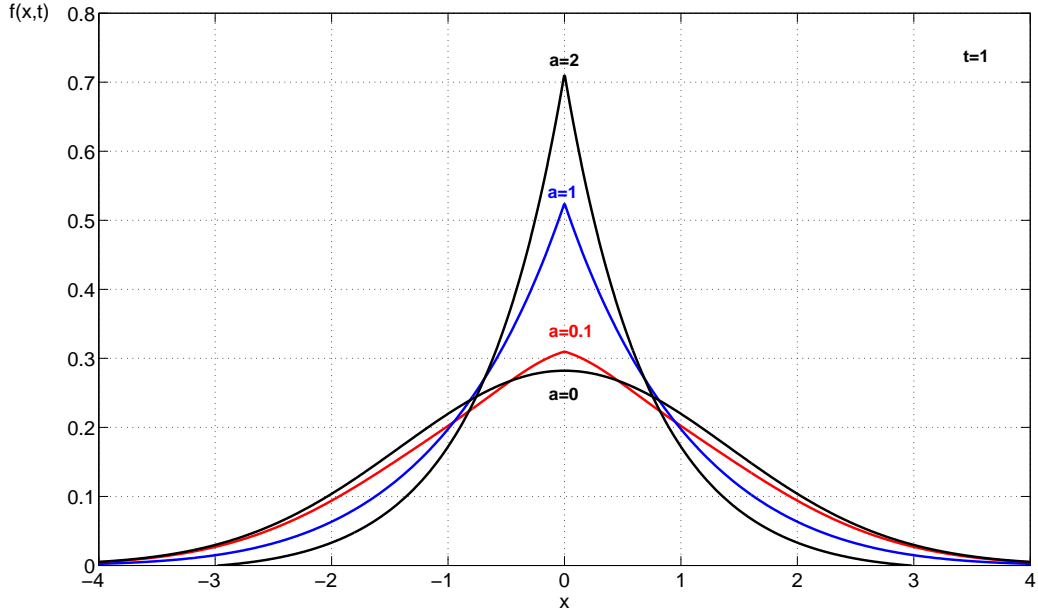}
\end{center}
\caption{Plot of the fundamental solution $f(x,t)$,  Eq. (\ref{brownber}), at time $t=1$, 
for different values of the parameter $a=[0,0.1,1,2]$. 
When $a=0$ we have the standard Gaussian density.}
\label{fig2Exp} 
\end{figure}
\noindent
{\bf Proof}: by Theorem \ref{t1} andEq. \ref{exptime2}), 
the fundamental solution of Eq. (\ref{expdiff}) is:
\begin{equation}
f(x,t)=\int_{0}^{\infty}G(x,\tau)f_{l}(\tau,t)\,d\tau
=e^{-at}G(x,t)+(1-e^{-at})\phi(x,t),
\label{brownber1}
\end{equation}
where 
$$
\phi(x,t)=\int_{0}^{\infty}G(x,t)\varphi(\tau,t)d\tau.
$$
We have that:
$$
\phi(x,t)=\frac{a}{1-e^{-at}}\int_{0}^{\infty}G(x,\tau)e^{-a\tau}\theta(t-\tau)d\tau.
$$
One has to evaluate:
\begin{equation}
\chi(x,t)=\int_{0}^{t}\frac{e^{-x^2/4\tau}e^{-a\tau}}{\sqrt{4\pi \tau}}d\tau,\;\; x\in \mathbb{R},t\ge 0.
\label{chiex}
\end{equation}
First we observe that:
$$
\chi(0,t)=\int_{0}^{t}\frac{e^{-a\tau}}{\sqrt{4\pi \tau}}d\tau=\frac{1}{2\sqrt a}\frac{2}{\sqrt \pi}\int_{0}^{\sqrt{at}}e^{-y^2}dy=\frac{1}{2\sqrt a}\textrm{ Erf}(\sqrt {at}).
$$
after the change of variables $y=\sqrt{a\tau}$. Because Erf$(-u)=-$Erf$(u)$ we can write:
\begin{equation}
\chi(0,t)=\frac{1}{4\sqrt a}\left\{\textrm{ Erf}(\sqrt{at})-\textrm{ Erf}(-\sqrt{at})\right\}.
\label{firso}
\end{equation}
Now, for any $x\in \mathbb{R}$:
\begin{equation}
\chi(x,t)=\frac{1}{4\sqrt a}\left\{e^{x\sqrt a}\textrm{ Erf}\left(\frac{x}{2\sqrt t}+\sqrt{at}\right)-e^{-x\sqrt a}\textrm{ Erf}\left(\frac{x}{2\sqrt t}-\sqrt{at}\right)  \right\}-\frac{1}{2\sqrt a}\sinh (|x|\sqrt a),
\label{chiex2}
\end{equation}
because:
$$
\frac{d}{d\tau}\left[\frac{1}{4\sqrt a}\left\{e^{x\sqrt a}\textrm{ Erf}\left(\frac{x}{2\sqrt \tau}+\sqrt{a\tau}\right)-e^{-x\sqrt a}\textrm{ Erf}\left(\frac{x}{2\sqrt \tau}-\sqrt{a\tau}\right)  \right\}\right]
$$
$$
=\frac{1}{4\sqrt a}\left\{\frac{2}{\sqrt \pi}e^{x\sqrt a}\exp\left(-\left[\frac{x}{2\sqrt \tau}+\sqrt{a\tau}\right]^2\right)\left(-\frac{x}{4}\tau^{-3/2}+\frac{\sqrt a}{2\sqrt \tau}\right)\right\}
$$
$$
-\frac{1}{4\sqrt a}\left\{\frac{2}{\sqrt \pi}e^{-x\sqrt a}\exp\left(-\left[\frac{x}{2\sqrt \tau}-\sqrt{a\tau}\right]^2\right)\left(-\frac{x}{4}\tau^{-3/2}-\frac{\sqrt a}{2\sqrt \tau}\right)  \right\}
$$
$$
=\frac{1}{\sqrt{4\pi \tau}}\exp\left(-\frac{x^2}{4\tau}-a\tau\right).
$$
Moreover, because Erf$(\pm\infty)=\pm 1$, we haveEq. \ref{chiex2}), 
which actually reduces to Eq. (\ref{firso}) when $x=0$. 
Therefore, the fundamental solution of Eq. (\ref{expdiff}) is:
\begin{equation}
f(x,t)=e^{-at}G(x,t)+\frac{\sqrt a}{4}\left\{e^{x\sqrt a}
\textrm{ Erf}\left(\frac{x}{2\sqrt t}+\sqrt{at}\right)-e^{-x\sqrt a}
\textrm{ Erf}\left(\frac{x}{2\sqrt t}-\sqrt{at}\right)  \right\}-\frac{\sqrt a}{2}\sinh (|x|\sqrt a),
\label{fonexp}
\end{equation}
which can be rewritten as Eq. (\ref{brownber}). $\Box$
\begin{rem}
With the choice (\ref{randexp}) the process:
$$
B(l(t))=B(b_tt+(1-b_t)j(t)),\;\; t\ge 0,
$$
has marginal density (\ref{brownber}). Observe that, for any $t\ge 0$:
$$
B(b_tt+(1-b_t)j(t))=^{\!\!\!\!{}^{d}}\;b_tB(t)+(1-b_t)B(j(t)),
$$
which naturally corresponds to Eq. \ref{brownber}).
\end{rem}
\begin{rem}
We observe that Eq. (\ref{brownber}) reduces to $G(x,t)$ when $a=0$ (see Figure \ref{fig2Exp}), 
which is as expected because the memory kernel disappears. 
For small times, the non-local memory effects are negligible and the process appears Markovian. 
Fig. 8 displays the fundamental solution at fixed $t$ and various values of $a$, 
whereas Fig. 9 displays the fundamental solution at fixed $a$ and various values of $t$.
\\ \\
For large times we have:  
$$
\lim_{t\rightarrow \infty}f(x,t)=\lim_{t\rightarrow \infty}\phi(x,t)=\overline\phi(x),
$$
where:
\begin{equation}
\overline \phi(x)=
\frac{\sqrt a}{2}(\cosh (x\sqrt a)-\sinh(|x|\sqrt a))=\frac{\sqrt{a}}{2}e^{-|x|\sqrt a},\;\; x\in \mathbb{R}.
\label{asymd}
\end{equation}
\end{rem}
\begin{figure}[!h]
\begin{center} 
\includegraphics[keepaspectratio=true,height=9cm]{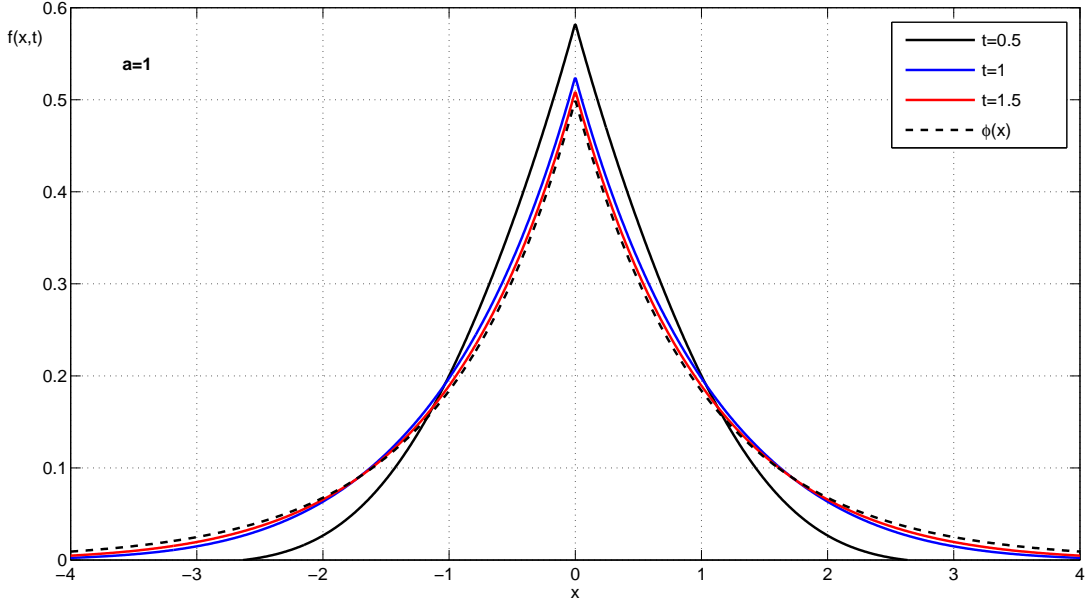}
\end{center}
\caption{Plot of the fundamental solution $f(x,t)$, Eq. (\ref{brownber}),
 at time $t=[0.5,1,1.5]$, and $a=1$. 
 The dashed line represents the asymptotic distribution $\overline \phi(x)$, Eq. (\ref{asymd}).} 
\end{figure}

\begin{rem}\label{remstat}
In view of Eq. (\ref{exptime}), 
it is always possible to choose the random time process $l(t)$, $t\ge 0$, 
such that it becomes  stationary at large times, 
in the sense of finite-dimensional densities. 
With this choice, the subordinated process $B(l(t))$ tends to a stationary process 
with asymptotic marginal distribution given by Eq. (\ref{asymd}). 
For instance, if we look at Eq. (\ref{overtime}), 
as $t \rightarrow \infty$ 
we have $\overline l(t)=X\sim ae^{-a\tau}$, $\tau \ge 0$.
A less trivial example can be constructed by replacing the random variable $X$ 
with a stationary process $X(t)$, $t\ge 0$, such that for each $t\ge 0$ the random variable $X(t)$ 
has an exponential distribution with mean $E(X(t))=a^{-1}$. 
The resulting process $l(t)=\min (X(t),t)$ 
is not increasing, has marginal distribution defined by 
Eq. (\ref{exptime}) and tends to $X(t)$ for large $t$. See Fig. 10.
\end{rem}
\begin{rem}
To obtain an idea on how fast the stationary regime is reached, one can look 
at the variance of the subordinated process. 
Using Eq. (\ref{explmom}) with $m=1$, we find:
\begin{equation}
E(B(l(t))^2)=\frac{2}{a}(1-e^{-at}),
\label{varrexp	}
\end{equation}
which, for large times, tends  exponentially to $2/a$ (i.e. the variance of eq.~\ref{asymd}).
\end{rem}
\begin{figure}[!h]
\begin{center} 
\includegraphics[keepaspectratio=true,height=10cm]{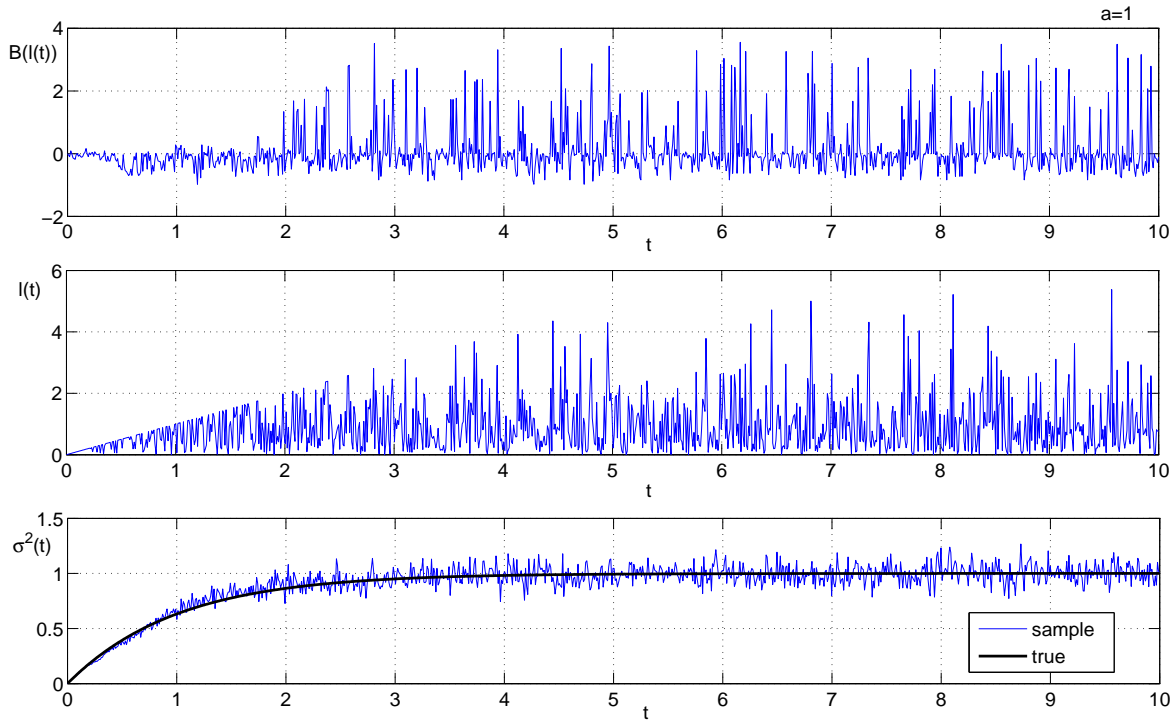}
\end{center}
\caption{Trajectory of the process $B(l(t))$ (top panel),
 with $0<t<10$, $E(B(1))=1$, $l(t)=\min(t,X(t))$ where $X(t)$ is 
 an exponential White Noise with mean one. 
 The corresponding trajectory of the random time $l(t)$ 
 process is presented in the middle panel. 
 The estimated variance is computed on a sample of dimension $N=500$. 
 The smooth black line in the bottom panel corresponds to $\sigma^2(t)$ given 
 by Eq. (\ref{varrexp }) and the stationary value is 
 $\displaystyle\lim_{t\rightarrow \infty}\sigma^2(t)=1$. \label{f4}} 
\end{figure}
\subsection{Exponential-decay kernel with logarithmic scaling time}
What happens if we choose an exponential kernel $K(t)=e^{-at}$ 
and a logarithmic scaling time? That is:
\begin{equation}
g(t)=\log(t+1),\;\; t\ge 0.
\end{equation}
Since $g'(t)K(g(t)-g(s))=(t+1)^{-a}(s+1)^{a-1}$, we get:
\begin{equation}
u(x,t)=u_0(x)+\frac{1}{(t+1)^a}\int_{0}^{t}(s+1)^{a-1}\partial_{xx}u(x,s)ds.
\end{equation}
Its fundamental solution is:
$$
f(x,t)=\frac{1}{(t+1)^a}G(x,\log(t+1))-\frac{\sqrt a}{2}\sinh (|x|\sqrt a)
$$
\begin{equation}
+\frac{\sqrt a}{4}\left\{e^{x\sqrt a}\textrm{ Erf}\left(\frac{x}{2\sqrt {\log(t+1)}}+\sqrt{a{\log(t+1)}}\right)-e^{-x\sqrt a}\textrm{ Erf}\left(\frac{x}{2\sqrt {\log(t+1)}}-\sqrt{a{\log(t+1)}}\right)  \right\}
\label{fonexplog}
\end{equation}
\begin{rem}
As in Remark \ref{remstat}, consider a random time process $l(t)$, $t\ge 0$, 
with marginal distribution defined by Eq. (\ref{exptime}), that becomes stationary  for large times. 
The subordinated process $B(l(\log(t+1)))$, $t\ge 0$, has  marginal density function defined by $f(x,t)$ 
of Eq. (\ref{fonexplog}). 
Observe that in this case the random time process $l(\log(t+1))$ 
is no longer asymptotically stationary. 
This is because the translational time-invariance is broken by the logarithmic transformation. 
However, we can always consider a random time process $l^{{}^*}(t)$, $t\ge 0$, 
with the same marginal distribution of $l\left(\log(t+1)\right)$, 
which becomes stationary for large times. 
Thus, the process $B(l^{{}^*}(t))$ still has a marginal density function defined by $f(x,t)$  
but becomes stationary as $t\rightarrow \infty$, 
in the sense of finite-dimensional distribution, with asymptotic marginal distribution given 
by Eq. (\ref{asymd}).  See Fig. 11.   
\end{rem} 
\begin{rem}
While $B(l(t))$, $t\ge 0$, satisfies Eq. (\ref{varrexp }) 
and thus has a variance which tends exponentially fast to the limit value $2/a$, 
here the stationary regime is reached more slowly. 
Indeed, the variance of the subordinated process is:
\begin{equation}
E(B(l^{{}^*}\!\!(t))^2)=\frac{2}{a}\left(1-\frac{1}{(t+1)^a}\right),
\label{varrexplog} 
\end{equation} 
which, for large times, converges to the stationary value $2/a$ with a power-like behavior.
\end{rem}
\begin{figure}[!h]
\begin{center} 
\includegraphics[keepaspectratio=true,height=10cm]{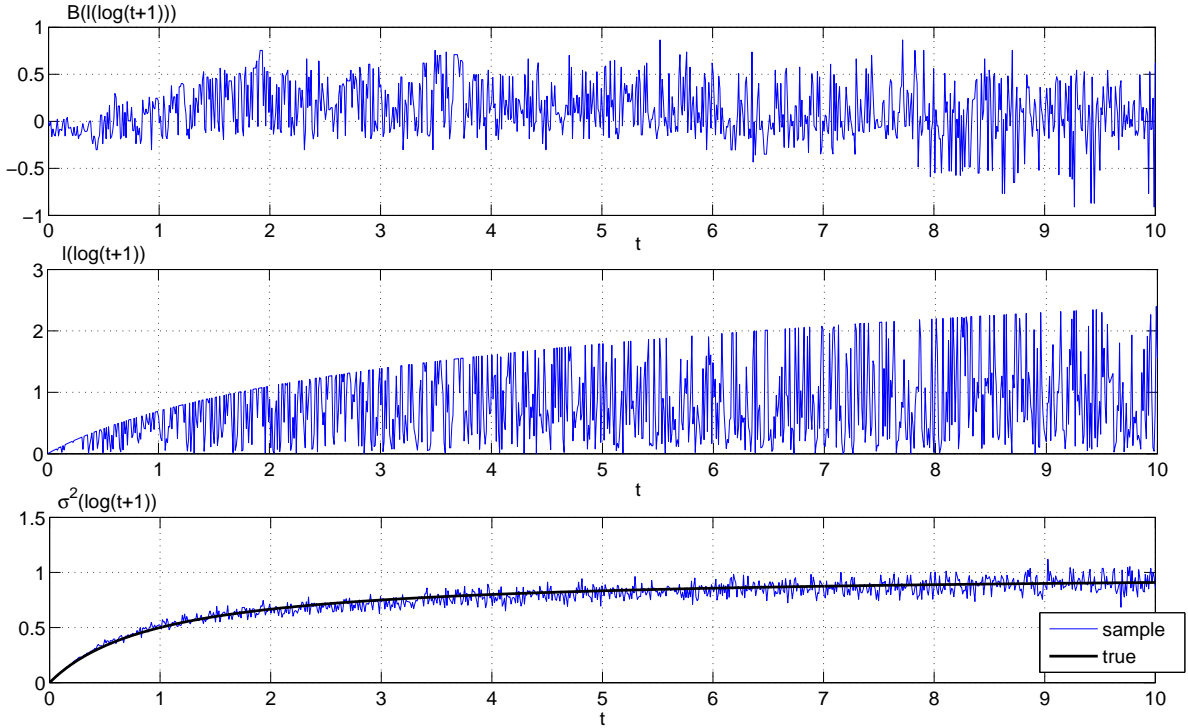}
\end{center}
\caption{Trajectory of the process $B(l(\log(t+1)))$ 
(top panel), with $0<t<10$, $E(B(1))=1$, $l(t)=\min(t,X(t))$ where $X(t)$ 
is an exponential White Noise with mean one. 
The corresponding trajectory of the random time process $l(\log(t+1))$ 
is presented in the middle panel. 
The estimated variance is computed on a sample of dimension $N=500$. 
The smooth black line in the bottom panel corresponds to Eq. (\ref{varrexp }). 
The stationary value is $\displaystyle\lim_{t\rightarrow \infty}\sigma^2(\log(t+1))=1$. 
The stationary regime is achieved more slowly than in the case of Figure \ref{f4}.} 
\end{figure}

\section{Examples involving other diffusions}\label{A}
We shall now consider examples of fractional and stretched Fokker-Planck equations 
involving  diffusion operators other than 
$\mathcal P_x=\partial_{xx}$ 
which corresponds to standard Brownian motion. 
We will choose $K(t)=t^{\beta-1}/\Gamma(\beta)$ and $g(t)=t^{\alpha/\beta}$ as in Section \ref{S4}.1 
and consider the partial integro-differential equation:
\begin{equation}
u(x,t)=u_0(x)+\frac{1}{\Gamma(\beta)}\frac{\alpha}{\beta}\int_{0}^{t} s^{\frac{\alpha}{\beta}-1}\left(t^{\frac{\alpha}{\beta}}-s^{\frac{\alpha}{\beta}}\right)^{\beta-1}\mathcal{P}_xu(x,s)ds,\;\;0<\beta\le 1,\;\; \alpha>0.
\label{tfpll}
\end{equation}
Its fundamental solution is the marginal density of the process:
\begin{equation}
\mathcal{D}(t)=Q(l_{\beta}(t^{\alpha/\beta})),	
\end{equation}
where $Q(t)$, $t\ge 0$, is the stochastic diffusion associated to $\mathcal{P}_x$ 
and $l_\beta(t)$, $t\ge 0$, is a suitable self-similar random time process. 
One has the following particular cases: \\

$\bullet$ When $\alpha=\beta$ and $0<\beta\le 1$, Eq. (\ref{tfpll}) 
becomes the {\it ``time-fractional'' Fokker-Planck equation}:
\begin{equation}
u(x,t)=u_0(x)+\frac{1}{\Gamma(\beta)}\int_{0}^{t} \left(t-s\right)^{\beta-1}\mathcal{P}_xu(x,s)ds,\;\;0<\beta\le 1,
\label{FrPl}
\end{equation}
whose fundamental solutions are the marginal distributions of the process:
$$
\mathcal{D}(t)=Q(l_{\beta}(t)),
$$
and are given by:
\begin{equation}
f_\mathcal{D}(x,t)=\int_{0}^{\infty}f_Q(x,\tau)f_{l_\beta}(\tau,t)d\tau,
\end{equation}
where 
\begin{equation}
f_{l_\beta}(\tau,t)=t^{-\beta}M_{\beta}(\tau t^{-\beta}),\;\; \tau,t \ge 0,\;\; 0<\beta\le 1
\label{rtpp}
\end{equation}
and $f_{Q}(x,t)$ is the probability density of $Q(t)$.\\ 

$\bullet$ When $\beta=1$ and $\alpha>0$ we get a  {\it ``time-stretched'' Fokker-Planck equation}:
\begin{equation}
u(x,t)=u_0(x)+\int_{0}^{t}\alpha s^{\alpha-1} \mathcal{P}_xu(x,s)ds,\;\;0<\beta\le 1.
\end{equation}
In this case $f_l(\tau,t)=\delta(\tau-t^{\alpha})$ and we get:
$$
f_\mathcal{D}(x,t)=f_{Q}(x,t^{\alpha}),
$$
which corresponds to the ``stretched'' diffusion:
$$
\mathcal{D}(t)=Q(t^{\alpha}),\;\;\ \alpha>0.
$$

$\bullet$ The case $\alpha=\beta=1$ is trivial and corresponds merely to the Markovian case where 
the equation is:
$$
u(x,t)=u_0(x)+\int_{0}^{t}\mathcal{P}_xu(x,s)ds
$$
whose fundamental solution is the density function of $D(t)=Q(t)$,
 namely the Markovian process. \\

In the following subsections we study the above equations under particular choices 
of the Fokker-Planck operator $\mathcal{P}_x$. 
In all the cases considered, 
we also give the results involving the exponential-decay kernel.
\subsection{Brownian motion with drift}\label{bmwdr}
Let $\mu \in \mathbb{R}$ and $\sigma^2 >0$ be given. 
Consider a linear diffusion $B^{(\mu)}=B^{(\mu,\sigma)}$ on $\mathbb{R}$ 
satisfying the stochastic differential equation:
\begin{equation}
dB^{(\mu)}(t)=\mu dt+\sigma dB(t),\;\; t\ge 0,
\end{equation} 
where $B(t)$, $t\ge 0$, is a ``standard'' Brownian motion. 
The process $B^{(\mu)}(t)$, $t\ge 0$, is called {\it Brownian motion with drift} $\mu$. 
It corresponds merely to a Brownian motion plus a drift term, namely:
\begin{equation}
B^{(\mu)}(t)=\mu t+\sigma B(t),\;\; t\ge 0.
\label{Bdrift}
\end{equation} 
The marginal density function of $B^{(\mu)}$(t), $t\ge 0$, is:
\begin{equation}
f_{B^{(\mu)}}(x,t)=\frac{1}{|\sigma| \sqrt{4\pi t}}\exp\left(-\frac{(x-\mu t)^2}{\sigma^24t} \right),\;\; t\ge 0,\;\; x\in
\mathbb{R},
\label{driftden}
\end{equation}
which is the fundamental solution of the Fokker-Planck equation:
\begin{equation}
\partial_tu(x,t)=-\mu \partial_xu(x,t)+\sigma ^2\partial_{xx}u(x,t),\;\; t\ge 0.
\end{equation}
\subsubsection{The $\beta$-power kernel.}
We consider the {\it ``fractional'' Fokker-Planck equation},
 see Eq. (\ref{FrPl}) (see also \cite{MetzlerKlafterPhysRep00}):
\begin{equation}
u(x,t)=u_0(x)+\frac{1}{\Gamma(\beta)}\int_{0}^{t} \left(t-s\right)^{\beta-1}(-\mu \partial_xu(x,s)+\sigma^2\partial_{xx}u(x,s))ds,\;\;0<\beta\le 1.
\label{DrfFl}
\end{equation}
Its fundamental solution can be regarded as the marginal density function of the process:
\begin{equation}	
D(t)=B^{(\mu)}(l_\beta(t)),\;\; t\ge 0,\;\; 0<\beta\le 1,
\label{drfitD}
\end{equation}
where the process $l_{\beta}(t)$, $t\ge 0$, 
is a self-similar random time process with parameter $H=\beta/2$, independent of $B^{(\mu)}$, 
such that its marginal distribution is given by Eq. (\ref{rtpp}). 
\begin{prop}
The fundamental solution of Eq. (\ref{DrfFl}) is:
$$
f_D(x,t)=\int_{0}^{\infty}f_{B^{(\mu)}}(x,\tau)f_{l_{\beta}}(\tau,t)\, d\tau,\;\; 
t\ge 0,\;\;\ x\in \mathbb{R},
$$
i.e.
\begin{equation}
f_D(x,t)=\int_{0}^{\infty}\frac{1}{|\sigma|\sqrt{4\pi \tau}}\exp\left(-\frac{(x-\mu \tau)^2}{4\sigma^2\tau} \right)\mathcal{M}_\beta(\tau,t)d\tau,\;\; t\ge 0,\;\; x\in \mathbb{R},
\label{intdrf}
\end{equation}
which is equal to:
\begin{equation}
f_D(x,t)=e^{\mu x/2\sigma^2}\frac{1}{2|\sigma|}\sum_{k=0}^{\infty}\frac{(-\mu^2t^{\beta}/4\sigma^2)^k}{k!}t^{-\beta/2}H^{2,0}_{2,2}\left(|x\sigma^{-1}|t^{-\beta/2}\Big{|}\begin{array}{c}(1/2,1/2), (1-\beta/2+\beta k,\beta/2)\\ (0,1),(k+1/2,1/2)\end{array}\right),
\label{phiH}
\end{equation}
where the Fox $H$-function is defined by Eq. (\ref{HMell}).
\end{prop}
\begin{figure}[!t]
\begin{center} 
\includegraphics[keepaspectratio=true,height=9cm]{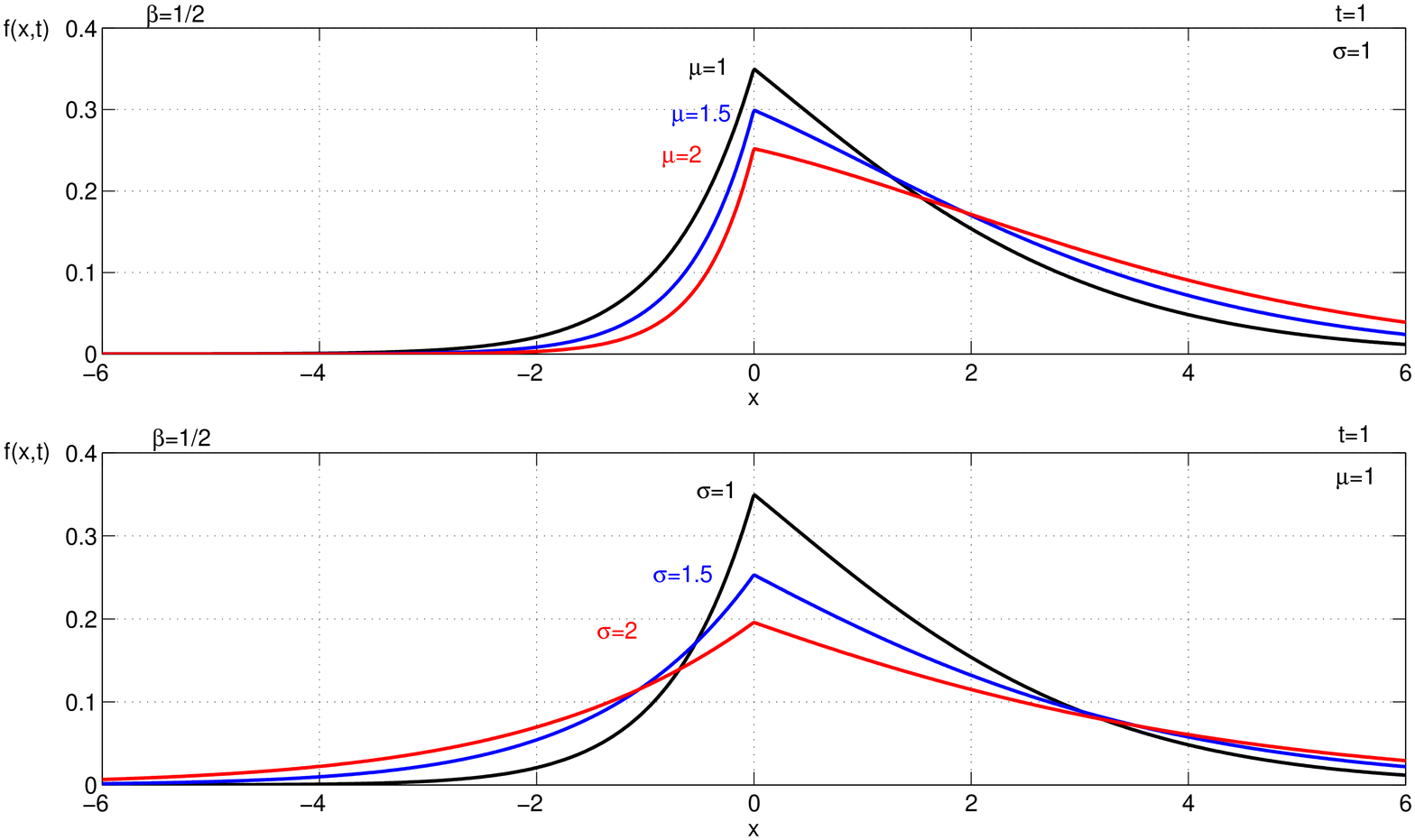}
\end{center}
\caption{Plot of the fundamental solution $f(x,t)$, 
Eq. (\ref{phiH}), with $\beta=1/2$, at time $t=1$, 
for different values of the parameters $\mu=[1,1.5,2]$ and $\sigma=[1,1.5,2]$. \label{drden}}
\end{figure}
\noindent
{\bf Proof}: In order to evaluate $f_D(x,t)$ we write:
$$
f_D(x,t)=
|\sigma|^{-1}e^{\mu x'/2\sigma}\int_{0}^{\infty}e^{-\mu^2\tau/4\sigma^2}G(x',\tau)
\mathcal{M}_\beta(\tau,t)d\tau,
$$
where $G(x,t)$ is the standard Gaussian density, see Eq. (\ref{ggauss}) 
and $x'=x/\sigma$. In view of Eq. (\ref{st3}), we have to evaluate an integral of the form:
\begin{equation}
\Phi(x,t)=\frac{1}{2}\int_{0}^{\infty}e^{-a\tau}\mathcal{M}_{1/2}(|x|,\tau)
\mathcal{M}_{\beta}(\tau,t)d\tau,\;\;\ x\in \mathbb{R},\;\;t\ge 0,\;\; a\ge 0.
\label{intof}
\end{equation}
One has:
$$
\Phi(x,t)=\frac{1}{2}\int_{0}^{\infty}e^{-a\tau}\tau^{-1/2} M_{1/2}(|x|\tau^{-1/2})t^{-\beta}
M_{\beta}(\tau t^{-\beta})\, d\tau
$$
$$
=\frac{1}{2}\int_{0}^{\infty}\frac{1}{y}M_{1/2}\left(\frac{|x|}{y}\right)2ye^{-ay^2}t^{-\beta}
M_{\beta}(y^2 t^{-\beta})\, dy.
$$
after  the change of variables $y=\sqrt \tau$. 
Because of the symmetry, it is enough to consider only the case $x \ge 0$. We get:
$$
\Phi(x,t)=\frac{1}{2}(M_{1/2}\star Y_t)(x),\;\;\ x\ge 0,
$$
where 
$$
(\varphi\star\phi)(x)=\int_{0}^{\infty}\frac{1}{y}\varphi\left(\frac{x}{y}\right)\phi(y)dy
$$
indicates the Mellin convolution and where:
\begin{equation}
Y_t(x)=2xe^{-ax^2}t^{-\beta}M_{\beta}(x^2 t^{-\beta}),\;\; x\ge 0, \;\;\; t\ge 0.
\end{equation}
\begin{figure}[!t]
\begin{center} 
\includegraphics[keepaspectratio=true,height=9cm]{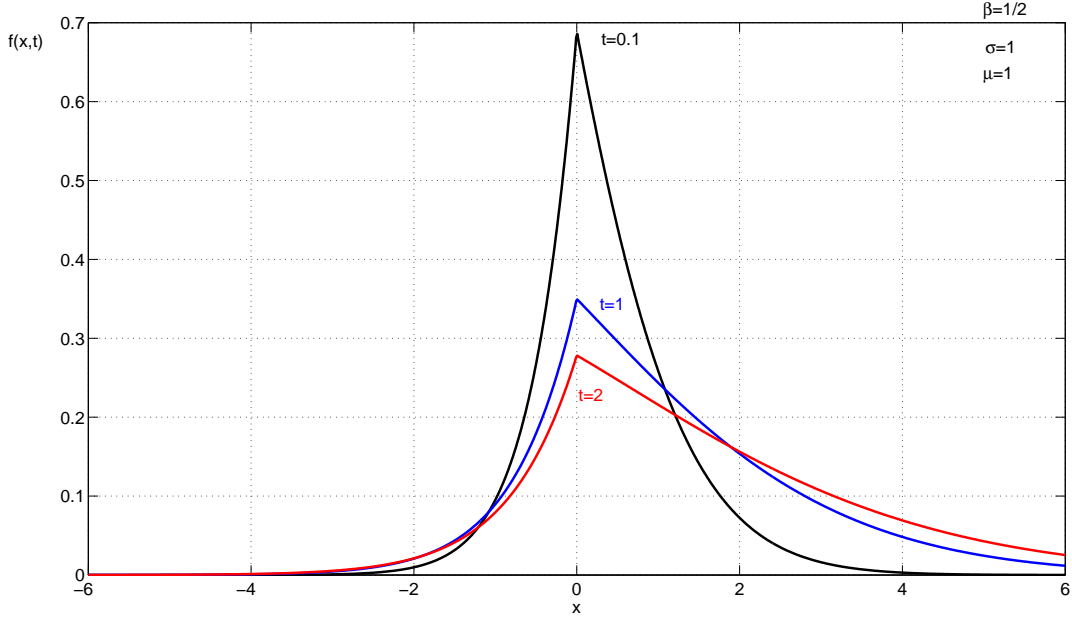}
\end{center}
\caption{Plot of the fundamental solution $f(x,t)$, Eq. (\ref{phiH}) with $\beta=1/2$,   
$\mu=1$, $\sigma=1$, at times $t=[0.1,1,2]$. \label{drden2}}
\end{figure}

Using the Mellin convolution theorem we get:
\begin{equation}
\mathscr{M}\{2\Phi(x,t);x,u\}=\mathscr{M}\{M_{1/2}(x);x,u\}\mathscr{M}\{Y_{t}(x);x,u\}.
\end{equation}
Because ofEq. \ref{Mfox}) andEq. \ref{HMell}), this can be written as:
\begin{equation}
\mathscr{M}\{2\Phi(x,t);x,u\}=\frac{\Gamma(u)}{\Gamma(1/2+u/2)}\mathscr{M}\{Y_{t}(x);x,u\}.
\end{equation}
We now evaluate:
$$
\mathscr{M}\{Y_{t}(x);x,u\}=\int_0^{\infty}e^{-ax^2}2xt^{-\beta}M_{\beta}(x^2t^{-\beta})x^{u-1}\,dx.
$$
After the change of variables $x^2t^{-\beta}=z$, we get
$$
\mathscr{M}\{Y_{t}(x);x,u\}=\int_{0}^{\infty}(zt^{\beta})^{\frac{1}{2}(u-1)}e^{-azt^{\beta}}M_\beta(z)dz
$$
$$
=t^{\frac{\beta}{2}(u-1)}\sum_{k=0}^{\infty}\frac{(-at^{\beta})^k}{k!}\int_{0}^{\infty}z^{k-\frac{1}{2}+\frac{u}{2}}M_\beta(z)dz
$$
$$
=t^{\frac{\beta}{2}(u-1)}\sum_{k=0}^{\infty}\frac{(-at^{\beta})^k}{k!}\mathscr{M}\{M_\beta(x);x,k+1/2+u/2\}
$$
$$
=t^{\frac{\beta}{2}(u-1)}\sum_{k=0}^{\infty}\frac{(-at^{\beta})^k}{k!}\frac{\Gamma(1/2+k+u/2)}{\Gamma(1+\beta k-\beta/2+\beta u/2)},
$$
where we have used Eq. (\ref{Mfox}) and Eq. (\ref{HMell}). Thus:
$$
\mathscr{M}\{2\Phi(x,t);x,u\}=
\sum_{k=0}^{\infty}\frac{(-at^{\beta})^k}{k!}t^{\frac{\beta}{2}(u-1)}\frac{\Gamma(u)\Gamma(1/2+k+u/2)}{\Gamma(1/2+u/2)\Gamma(1+\beta k-\beta/2+\beta u/2)}.
$$
\begin{figure}[!t]
\begin{center} 
\includegraphics[keepaspectratio=true,height=10cm]{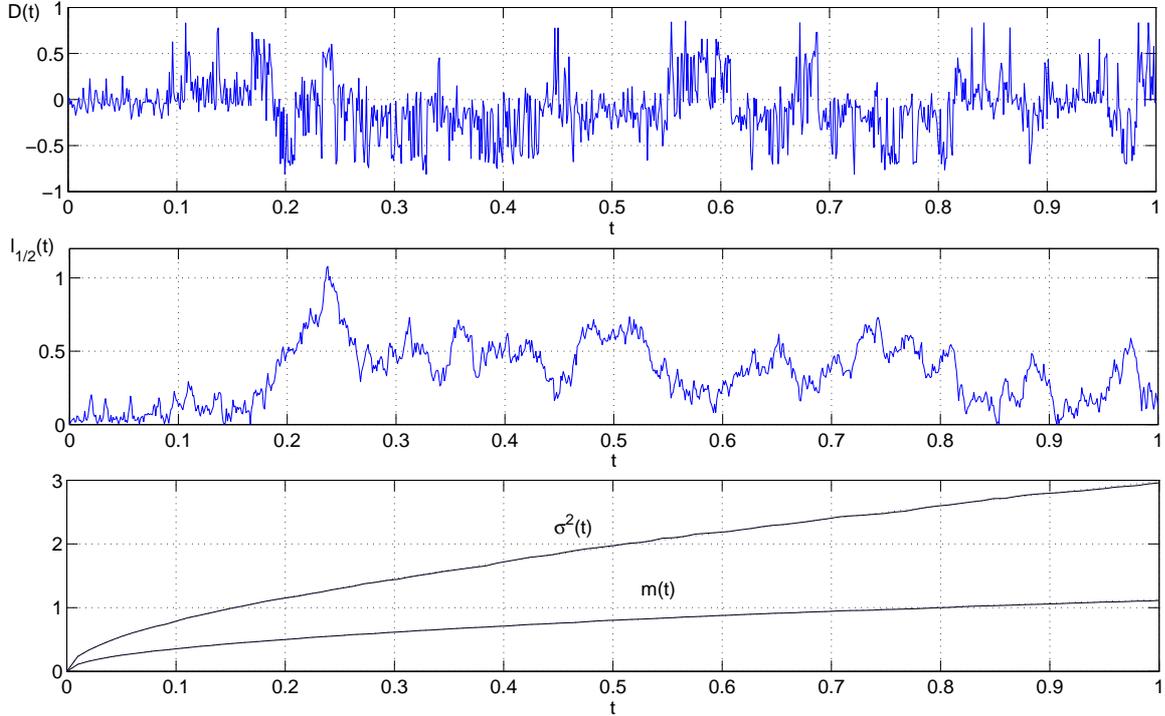}
\end{center}
\caption{Trajectory of the process $D(t)=B^{(\mu)}(l_{1/2}(t))$ 
defined in Eq. \ref{drfitD}) with $\beta=1/2$ (top panel). 
The random time process is $l_{1/2}(t)=|b(t)|$, where $b(t)$ 
is a ``standard'' Brownian motion (middle panel). 
The variance and the mean are evaluated over a sample of size $N=5\cdot 10^4$ 
and fit the theoretical values (bottom panel).\label{drtr}}
\end{figure}
Inverting the Mellin transform,Eq. \ref{HMell}) gives:
\begin{equation}
\Phi(x,t)=\frac{1}{2}\sum_{k=0}^{\infty}\frac{(-at^{\beta})^k}{k!}t^{-\beta/2}H^{2,0}_{2,2}\left(|x|t^{-\beta/2}\Big{|}\begin{array}{c}(1/2,1/2), (1-\beta/2+\beta k,\beta/2)\\ (0,1),(k+1/2,1/2)\end{array}\right),
\end{equation}
with $x\in \mathbb{R}$ and $t\ge 0$. 
Therefore, the fundamental solution of Eq. (\ref{FrPl}) can be expressed as:
$$
f_D(x,t)=e^{\mu x/2\sigma^2}\frac{1}{2|\sigma|}\sum_{k=0}^{\infty}\frac{(-\mu^2t^{\beta}/4\sigma^2)^k}{k!}t^{-\beta/2}H^{2,0}_{2,2}\left(|x\sigma^{-1}|t^{-\beta/2}\Big{|}\begin{array}{c}(1/2,1/2), (1-\beta/2+\beta k,\beta/2)\\ (0,1),(k+1/2,1/2)\end{array}\right),
$$
that is Eq. (\ref{phiH}). $\Box$\\

When $\mu=0$ and $\sigma=1$,Eq. \ref{phiH}) reduces to:
$$
f_D(x,t)=\frac{1}{2}t^{-\beta/2}H^{2,0}_{2,2}\left(|x|t^{-\beta/2}\Big{|}\begin{array}{c}(1/2,1/2), (1-\beta/2,\beta/2)\\ (0,1),(1/2,1/2)\end{array}\right),
$$
that is, using the reduction formula for the Fox $H$-function \cite{mainardiH},
$$
 f_D(x,t)=\frac{1}{2}t^{-\beta/2}H^{1,0}_{1,1}
 \left(|x|t^{-\beta/2}\Big{|}\begin{array}{c}(1-\beta/2,\beta/2)\\ (0,1)\end{array}\right)=
 \frac{1}{2}\mathcal{M}_{\beta/2}(|x|,t).
$$
As expected, we recover in this case the fundamental solution of the time-fractional diffusion 
equation  (\ref{gbm}).\\

Moreover, if we set $\beta=1$ in Eq. (\ref{phiH}) we have:
$$
f_D(x,t)=e^{\mu x/2\sigma^2}\frac{1}{2|\sigma|}
\sum_{k=0}^{\infty}\frac{(-\mu^2t/4\sigma^2)^k}{k!}t^{-1/2}H^{2,0}_{2,2}\left(|x\sigma^{-1}|t^{-1/2}\Big{|}
\begin{array}{c}(1/2,1/2), (1/2+k,1/2)\\ (0,1),(1/2+k,1/2)\end{array}\right),
$$
$$
=e^{\mu x/2\sigma^2}\sum_{k=0}^{\infty}\frac{(-\mu^2t/4\sigma^2)^k}{k!}\frac{1}{2|\sigma|}
t^{-1/2}H^{1,0}_{1,1}\left(|x\sigma^{-1}|t^{-1/2}\Big{|}\begin{array}{c}(1/2,1/2)\\ (0,1)\end{array}\right)=\frac{1}{|\sigma|\sqrt{4\pi t}}\exp\left(-\frac{(x-\mu t)^2}{4\sigma^2t} \right),
$$
and we recover $f_{B^{(\mu)}}(x,t)$.\\ 
\begin{figure}[!t]
\begin{center} 
\includegraphics[keepaspectratio=true,height=9cm]{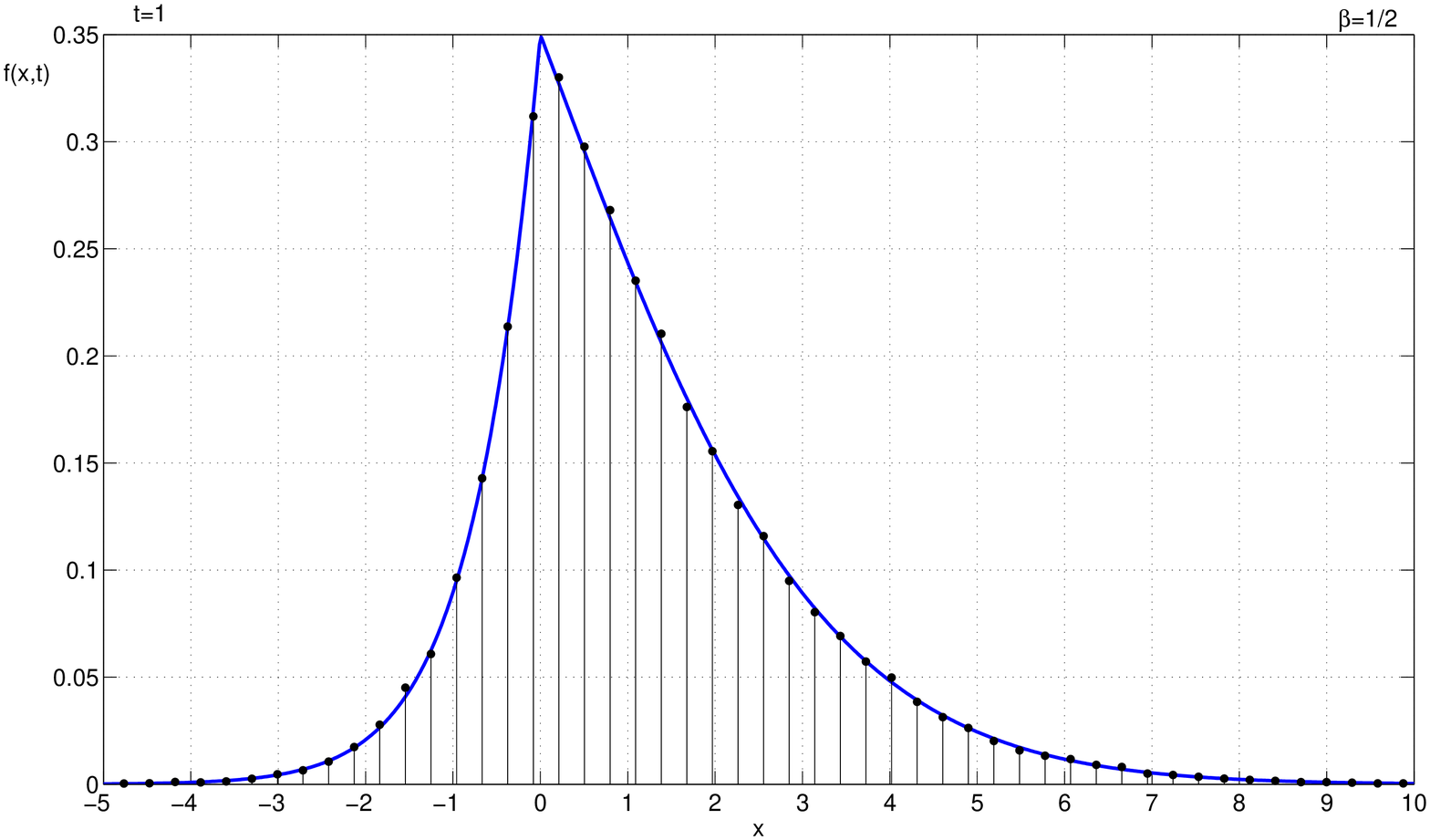}
\end{center}
\caption{Marginal density function $f(x,t)$ 
of the process $B^{(\mu)}(l_{1/2}(t))$ at time $t=1$Eq. \ref{phiH}). 
The histogram is evaluated over $N=10^5$ simulated trajectories (see Figure \ref{drtr}).\label{drifthist}}
\end{figure}

In Figure \ref{drden} and Figure \ref{drden2} we have 
used Eq. (\ref{intdrf}) to plot the fundamental solution (\ref{phiH})
 with $\beta=1/2$ for different values of the parameters $\mu$ and $\sigma$ at fixed time 
 and, for fixed parameters, at different times $t$. 
 As expected, the fundamental solution is not symmetric in space with a time-growing skewness. 
 Moreover, due to the presence of the positively taken drift term ($\mu=1$), 
 the probability to find the particle in the positive semi-axis increases with time (fig. \ref{drden2}).\\ 

In Figure \ref{drtr} is presented a trajectory of the process $D(t)=B^{(\mu)}(l_\beta(t))$ 
with $\beta=1/2$. Using Eq. (\ref{lmom}) it is easy to write all the moments of the process:
\begin{equation}
E(D(t)^m)=\sum_{k=0}^{[m/2]} 
\left(\!\!\begin{array}{cc} m\\ 2k\end{array}\!\!\right) \frac{2k!(m-k)!}{k!}\sigma^{2k}\mu^{m-2k}
\frac{t^{\beta(m-k)}}{\Gamma(\beta(m-k)+1)},\;\; 0<\beta \le 1,
\end{equation}
where $m$ is an integer greater than zero and $[ a ]$ 
indicates the integer part of $a$. Therefore, we have:
\begin{equation}
m(t)=E(D(t))=\mu \frac{t^\beta}{\Gamma(\beta +1)},
\end{equation}
and	
\begin{equation}
\sigma^2(t)=
E(D(t)^2)-m(t)^2=2\mu^2\frac{t^{2\beta}}{\Gamma(2\beta+1)}-
\mu^2\frac{t^{2\beta}}{\Gamma(\beta+1)^2}+2\sigma^2\frac{t^\beta}{\Gamma(\beta+1)}.
\end{equation}
In the bottom panel of Figure \ref{drtr} the mean and the variance 
have been estimated from a sample of trajectories of the process $B^{(\mu)}(l_{1/2})(t)$. 
Then, they have been compared with the theoretical values given above. In Figure \ref{drifthist} 
we compare the theoretical density function $f(x,t)$ given by Eq. \ref{phiH}) at time $t=1$ 
with an histogram evaluated over a sample of $N=10^5$ trajectories.   
\subsubsection{Exponential-decay kernel}
The {\it exponential-decay kernel} case is straightforward. 
The non-Markovian Fokker-Planck equation is:
\begin{equation}
u(x,t)=u_0(x)+\int_{0}^{t} e^{-a(t-s)}(-\mu \partial_xu(x,s)+\sigma^2\partial_{xx}u(x,s))ds,\;\;a\ge 0.
\label{DrfFlexp}
\end{equation}
If we indicate by $\mathcal{G}(x,t)$ the fundamental solution of the Markovian equation; i.e.
Eq. (\ref{driftden})
\begin{equation}
\mathcal{G}(x,t)=\frac{1}{|\sigma|\sqrt{4\pi t}}\exp\left(-\frac{(x-\mu t)^2}{4\sigma^2t} \right),\;\; t\ge 0,\;\; x\in
\mathbb{R},
\end{equation}
then, using Eq. (\ref{exptime}), the fundamental solution of Eq. (\ref{DrfFlexp}) is:
\begin{equation}
f(x,t)=e^{-at}\mathcal{G}(x,t)+(1-e^{-at})\Phi(x,t),
\end{equation}
where:
$$
\Phi(x,t)=
\frac{a}{1-e^{-at}}e^{\mu x/2\sigma^2}\int_{0}^{t}\frac{e^{-x^2/4\sigma^2\tau}
e^{-(a+\mu^2/4\sigma^2)\tau}}{|\sigma|\sqrt{4\pi \tau}}\, d\tau.
$$
Using Eq. (\ref{chiex}) and Eq. (\ref{chiex2}) we have:
\begin{prop}
The fundamental solution of Eq. (\ref{DrfFlexp}) is:
$$
f(x,t)=e^{-at}\mathcal{G}(x,t)-
\frac{ae^{\frac{\mu}{2\sigma^2} x}}{2|\sigma|\sqrt{a+\frac{\mu^2}{4\sigma^2}}}
\sinh \left(|x\sigma^{-1}|\sqrt {a+\frac{\mu^2}{4\sigma^2}}\right)+
$$
$$
+\frac{ae^{\frac{\mu}{2\sigma^2}x}}{4|\sigma|
\sqrt{a+\frac{\mu^2}{4\sigma^2}}}\left\{\exp\left(\frac{x}{|\sigma|}\sqrt {a+\frac{\mu^2}{4\sigma^2}}
\right)\textrm{\rm Erf}\left(\frac{x}{2|\sigma|\sqrt t}+
\sqrt{\Big{(}a+\frac{\mu^2}{4\sigma^2}\Big{)}t}\right)\right.
$$
\begin{equation}
\left.-\exp\left(-\frac{x}{|\sigma|}\sqrt {a+\frac{\mu^2}{4\sigma^2}}\right)
\textrm{\rm Erf}\left(\frac{x}{2|\sigma|\sqrt t}-\sqrt{\Big{(}a+\frac{\mu^2}{4\sigma^2}\Big{)}t}\right)  \right\}.
\end{equation}
\end{prop}
\noindent
When $t\rightarrow \infty$ we obtain the stationary distribution:
$$
\overline \phi(x)=\frac{ae^{\frac{\mu}{2\sigma^2} x}}{2|\sigma|\sqrt{a+\frac{\mu^2}{4\sigma^2}}}
\left(\cosh \left(x|\sigma^{-1}|\sqrt {a+\frac{\mu^2}{4\sigma^2}}\right)-\sinh
\left(|x\sigma^{-1}|\sqrt {a+\frac{\mu^2}{4\sigma^2}}\right)\right)
$$
that is:
\begin{equation}
\overline \phi(x)=\frac{a}{2|\sigma|
\sqrt{a+\frac{\mu^2}{4\sigma^2}}}\exp\left(\mu x/2\sigma^2-|x\sigma^{-1}|\sqrt {a+\frac{\mu^2}{4\sigma^2}}\right).
\label{asymd2}
\end{equation}
\subsection{Geometric Brownian motion}\label{gmbrw}
Let $\mu \in \mathbb{R}$ and $\sigma^2>0$ be given. 
Consider a linear diffusion $S$ on $\mathbb{R}$ defined by the stochastic differential equation:
\begin{equation}
dS(t)=\mu S(t)dt+\sigma S(t)dB(t),\;\;t\ge 0,
\label{geomst}
\end{equation} 
where $B(t)$, $t\ge 0$, is a ``standard'' Brownian motion. 
The process $S(t)$, $t\ge 0$, is called {\it Geometric Brownian motion}. 
If $S$ starts in $x_0$ at time $t=0$ (i.e. $P(S(0)=x_0)=1$), 
then a solution of Eq. (\ref{geomst}) is:
\begin{equation}
S(t)=x_0\exp\left[(\mu-\sigma^2/2)t+\sigma B(t)\right],\;\; t\ge 0,\;\; x_0>0.
\label{gm}
\end{equation}
The marginal density function of $S(t)$ is the log-normal distribution:
\begin{equation}
f_S(x,t)=
\frac{1}{x|\sigma|\sqrt{4\pi t}}
\exp\left(-\frac{\Big{(}\log(x/x_0)-(\mu-\sigma^2/2)t\Big{)}^2}{\sigma^24t}\right),\;\;t\ge 0,\;\, x\ge 0.
\label{logn}
\end{equation}
The function $f_S(x,t)$ is a solution of the Fokker-Planck equation:
\begin{equation}
\partial_t u(x,t)=\left[(2\sigma^2-\mu)+(4\sigma^2-\mu)x\partial_x+\sigma^2 x^2 \partial_{xx}\right]u(x,t),\;\; x\ge 0,
\label{geomfp}
\end{equation} 
with deterministic initial condition 
\begin{equation}
u_0(x)=\delta(x-x_0), \;\;\ x\ge 0,\;\; x_0>0.
\label{init} 
\end{equation}
\subsubsection{$\beta$-power kernel}
If we introduce the $\beta$-power kernel $K(t)=\Gamma(\beta)^{-1}t^{\beta-1}$, $0<\beta\le 1$, 
in this setting we obtain the following {\it ``fractional'' Fokker-Planck} equation:
\begin{equation}
u(x,t)=u_0(x)+
\frac{1}{\Gamma(\beta)}\int_{0}^{t} \left(t-s\right)^{\beta-1}\left[(2\sigma^2-\mu)+
(4\sigma^2-\mu)x\partial_x+\sigma^2 x^2 \partial_{xx}\right]u(x,s)ds,\;\; x\ge 0.
\label{GeomFl}
\end{equation}
A solution of the above equation with initial condition given by
Eq. (\ref{init}) is given by (see Corollary \ref{cor2}):
\begin{equation}
f_D(x,t)=\int_{0}^{\infty}f_S(x,\tau)\mathcal{M}_\beta(\tau,t)d\tau
\end{equation}
which is the marginal distribution of the process
\begin{equation}
D(t)=S(l_\beta(t)),\;\;t\ge 0,\;\; 0<\beta\le 1,
\label{fracge}
\end{equation}
starting almost surely in $x_0>0$, where 
$l_{\beta}(t)$, $t\ge 0$, 
is a self-similar random time process with $H=\beta/2$, 
independent of the geometric Brownian motion $S(t)$ 
and with marginal density function given by Eq. (\ref{rtpp}). 
It is easy to see that:
$$
f_D(x,t)=
\frac{1}{x|\sigma|}\exp{\left(\frac{\log(x/x_0)(\mu-\sigma^2/2)}{4\sigma^2}\right)}
\frac{1}{2}\int_{0}^{\infty}e^{-a\tau}\mathcal{M}_{1/2}(|x'|,\tau)\mathcal{M}_{\beta}(\tau,t)\, d\tau,
$$
where:
$$
a=(\mu-\sigma^2/2)^2/4\sigma^2,\;\; x'=\log(x/x_0)/\sigma.
$$
We have the same integral as in Eq. (\ref{intof}). Therefore:
\begin{prop}  for each $t\ge 0$:
$$
f_D(x,t)=\frac{1}{x|\sigma|}\exp\left(\frac{\log(x/x_0)(\mu-\sigma^2/2)}{4\sigma^2}\right)\times
$$
\begin{equation} 
\times \sum_{k=0}^{\infty}\frac{1}{k!}
\left(-\frac{(\mu-\sigma^2/2)^2t^{\beta}}{4\sigma^2}\right)^kt^{-\frac{\beta}{2}}H^{2,0}_{2,2}
\left(|x'|t^{-\frac{\beta}{2}}\Big{|}\begin{array}{c}(1/2,1/2), (1-\beta/2+\beta k,\beta/2)\\ 
(0,1),(k+1/2,1/2)\end{array}\right).
\label{phigH}
\end{equation}
\end{prop}
\begin{figure}[!t]
\begin{center} 
\includegraphics[keepaspectratio=true,height=9cm]{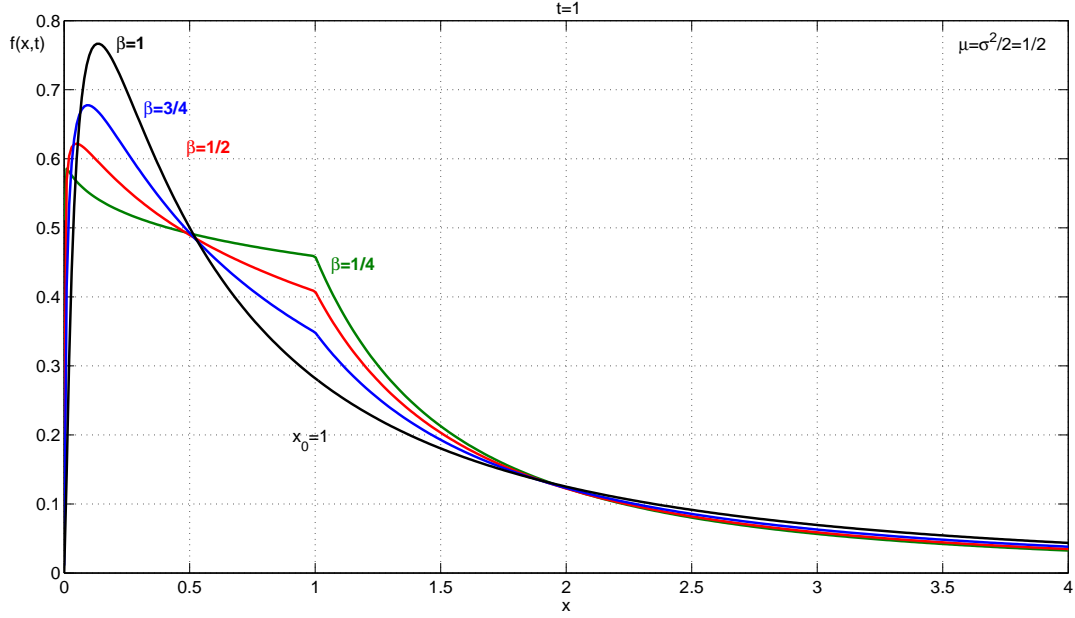}
\end{center}
\caption{Plot of the fundamental solution $f(x,t)$, Eq. (\ref{phigH}) at time $t=1$, 
when $\mu=\sigma^2/2=1$Eq. \ref{geompart}), $x_0=1$, 
for different values of the parameter $\beta=[1/4,1/2,3/4,1]$. 
For $\beta=1$ $f(x,t)$  reduces to the log-normal density (\ref{logn}). 
The angular point corresponds to the initial value $x_0=1$ and is due to the presence of 
$|\log(x/x_0)|$ in the solution. \label{geomden2}}
\end{figure}

This result can be obtained directly from
Eq. (\ref{phiH}) because our process is:
$$
D(t)=x_0\exp(B^{(\mu')}(l_\beta(t))),
$$
where $B^{(\mu')}$ is a Brownian motion with drift $\mu'=(\mu-\sigma^2/2)$. 
When $\beta =1$ we recover Eq. (\ref{logn}). 
Moreover, if $\mu=\sigma^2/2$ (i.e. $\mu'=0$) we have (see Figure \ref{geomden2}):
\begin{equation}
f_D(x,t)=\frac{1}{x|\sigma|}t^{-\beta/2}M_{\beta/2}\left(\bigg{|}\frac{\log(x/x_0)}{\sigma}\bigg{|}t^{-\beta/2}\right),\;\; x\ge 0,\;\;t\ge 0,
\label{geompart}
\end{equation}
which is the marginal probability density of:
$$
D(t)=x_0e^{\sigma B(l_\beta(t))},\;\, t\ge 0.
$$
\begin{figure}[!t]
\begin{center} 
\includegraphics[keepaspectratio=true,height=9cm]{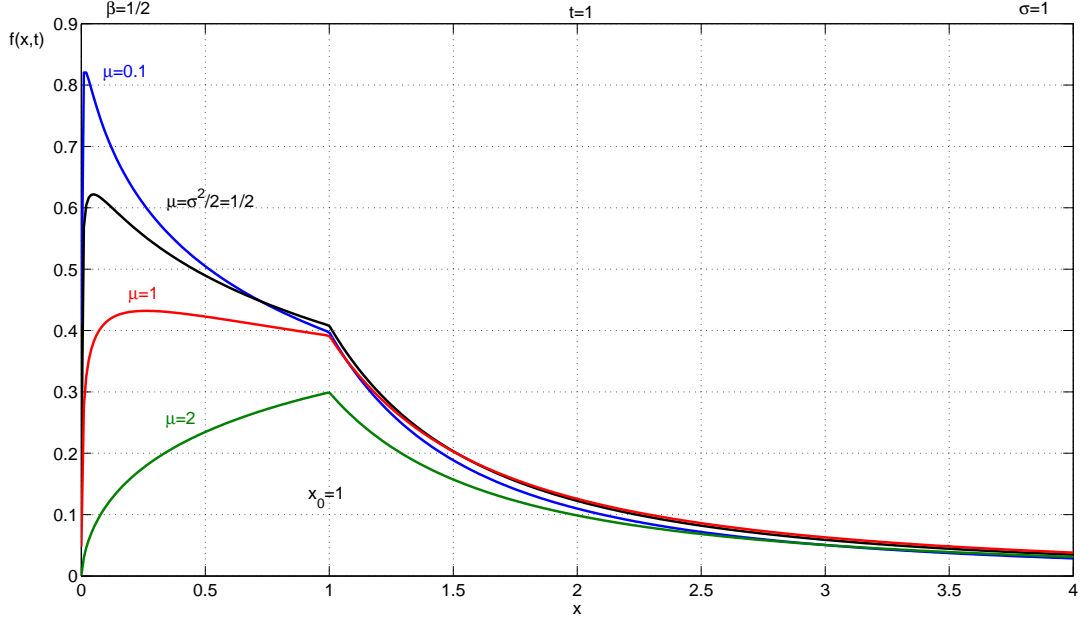}
\end{center}
\caption{Plot of the fundamental solution $f(x,t)$, Eq. (\ref{phigH}), with $\beta=1/2$, 
$\sigma=1$, $x_0=1$, at time $t=1$,  for different values of the parameter $\mu=[0.1,1/2,1,2]$. 
For $\mu=1/2$ we have Eq. (\ref{geompart}), 
see also Figure \ref{geomden2}.\label{geomden3}}
\end{figure}
\begin{figure}[!t]
\begin{center} 
\includegraphics[keepaspectratio=true,height=9cm]{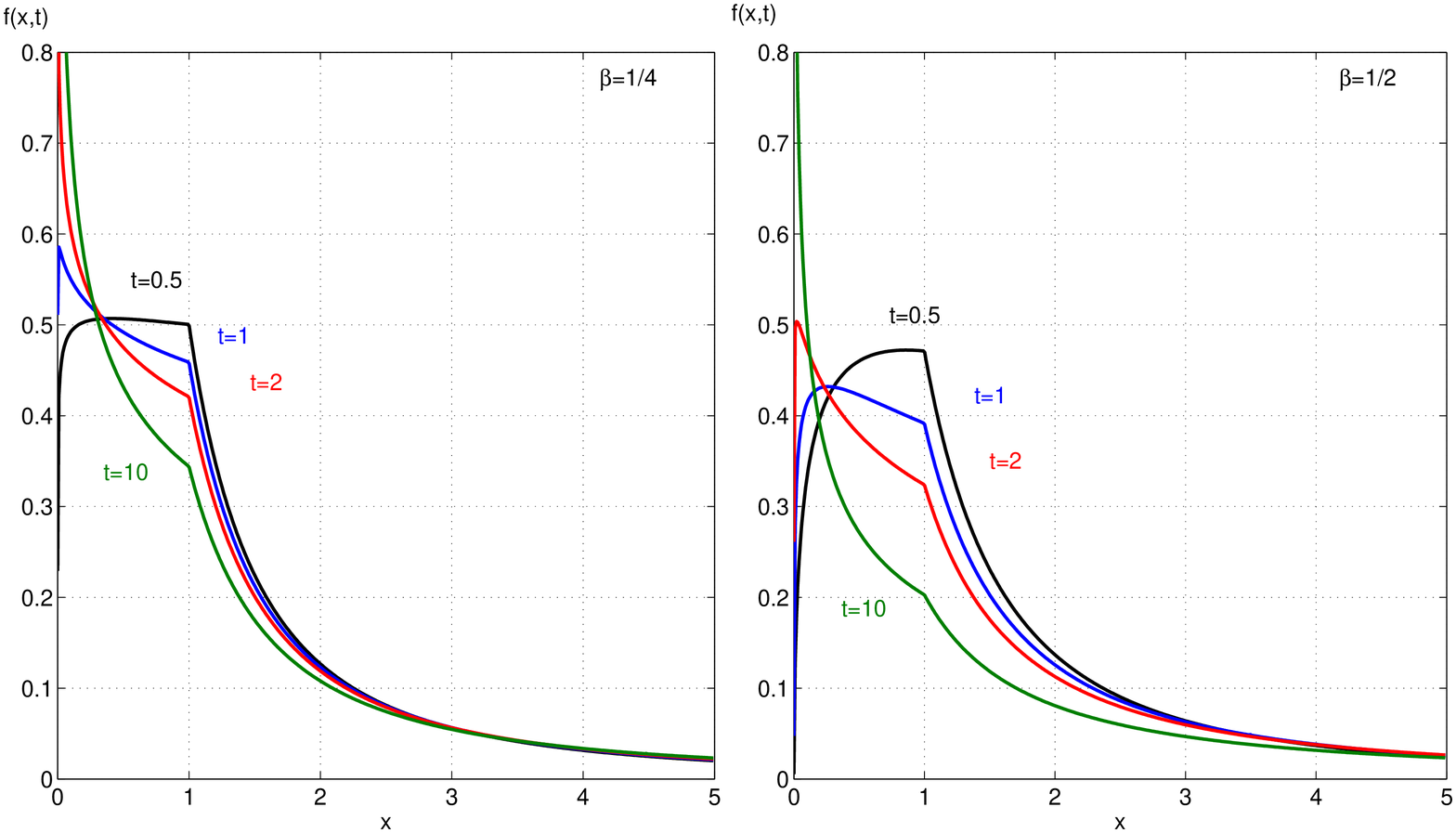}
\end{center}
\caption{Plot of the fundamental solution $f(x,t)$, Eq. (\ref{phigH}),
 with $\beta=1/4$ (left panel) and $\beta=1/2$ (right panel), at different times $t=[0.5,1,2,10]$.
 \label{geomden4}}
\end{figure}
In Figure \ref{geomden2} 
we show the plot of the fundamental solution $f(x,t)$ in the particular case given 
by Eq. (\ref{geompart}). 
Here we can see the behavior of the solution varying the parameter $\beta$. 
For $\beta=1$ we recover the log-normal density (\ref{logn}) with $\mu=\sigma^2/2$. 
In Figure \ref{geomden3} we point out the dependence of the solution with respect to the drift parameter 
$\mu$ for fixed $\beta=1/2$, $t=1$, $\sigma=1$ and $x_0=1$. 
In Figure \ref{geomden4} we present the time evolution of the fundamental solution with $\beta=1/2$ 
and $\beta=1/4$.\\

\begin{figure}[!t]
\begin{center} 
\includegraphics[keepaspectratio=true,height=10cm]{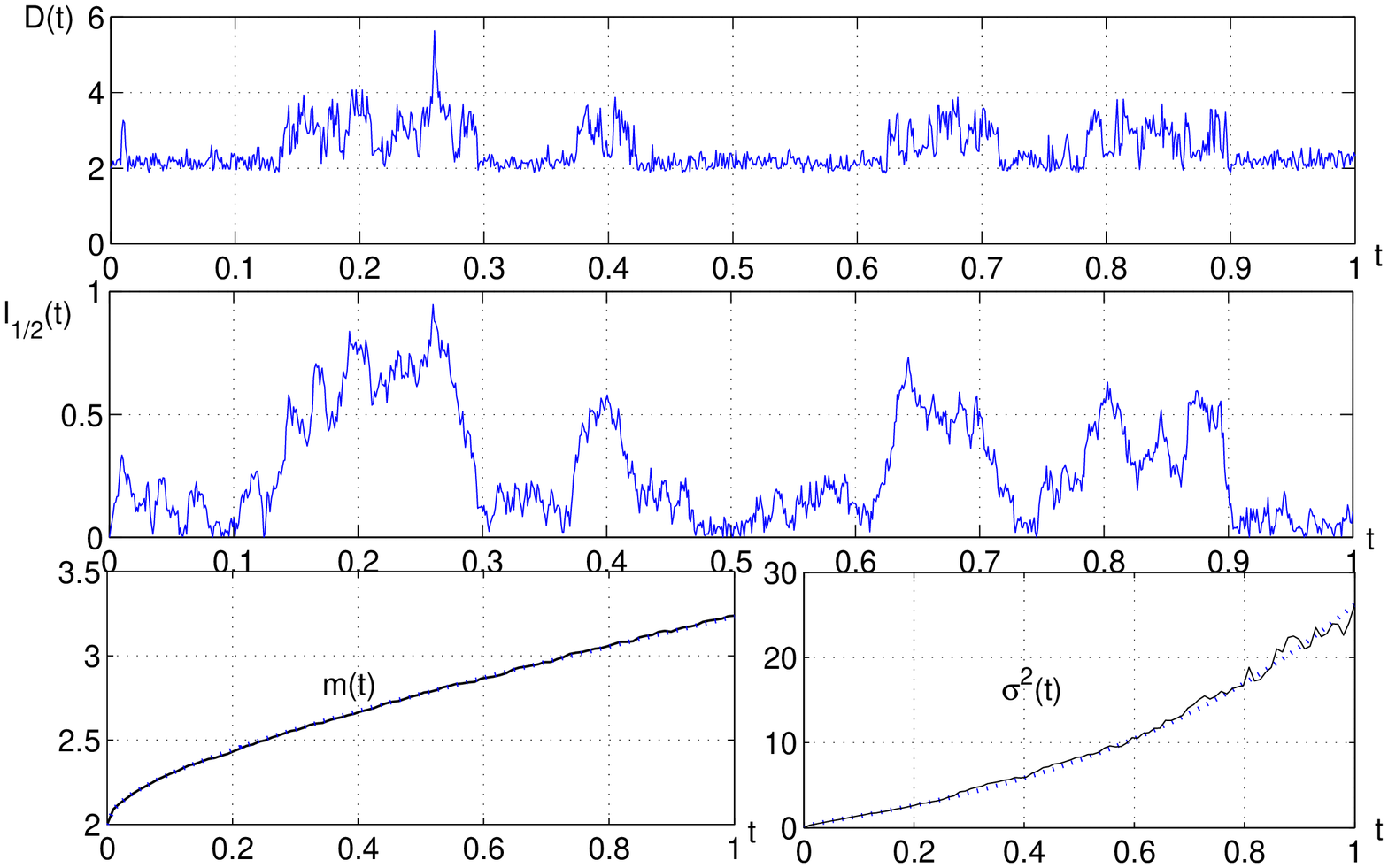}
\end{center}
\caption{Trajectory of the process $D(t)=S(l_{1/2}(t))$ defined 
in Eq. (\ref{fracge}) with $\beta=1/2$ (top panel). 
The random time process is $l_{1/2}(t)=|b(t)|$ (middle panel). 
The variance and the mean are evaluated over a sample of size $N=5\cdot 10^4$ 
and are presented together with the theoretical functions, Eq. (\ref{geommom}),
 in the bottom panels.\label{gmtr}}
\end{figure}

In Figure \ref{gmtr} we present a trajectory of the process 
$D(t)=S(l_{\beta}(t))$ with $\beta=1/2$, Eq. (\ref{fracge}). 
We shall now compute the mean and the variance of the process $D(t)$. We have that:
$$
E(S(t))=E\left (x_0\exp\left[(\mu-\sigma^2/2)t+\sigma B(t))\right] \right)=x_0\exp\left[(\mu-\sigma^2/2)t\right]E\left(e^{\sigma B(t)}\right).
$$
Therefore, because:
$$
E(e^{\sigma B(t)})=\int_\mathbb{R} e^{\sigma x}G(x,t)dx=
e^{\sigma^2 t}\frac{1}{\sqrt{4\pi t}}\int_{\mathbb{R}}e^{-\frac{(x-2\sigma t)^2}{4t}}\, dx,
$$
we have:
\begin{equation}
E(S(t))=x_0\exp\left[(\mu+\sigma^2/2)t\right].
\end{equation}
In the same way one has:
\begin{equation}
E(S(t)^2)=x_0^2\exp\left[(2\mu+3\sigma^2)t\right].
\end{equation}
Using the above equations we have:
$$
E(S(l_\beta(t)))=x_0E\left(\exp\left[(\mu+\sigma^2/2)l_\beta(t)\right]\right)=
x_0\sum_{k=0}^{\infty}\frac{(\mu+\sigma^2/2)^k}{k!}E(l_\beta(t)^k),
$$
which, using Eq. (\ref{lmom}), becomes:
$$
E(S(l_\beta(t)))=x_0\sum_{k=0}^{\infty}\frac{\left((\mu+\sigma^2/2)t^{\beta}\right)^k}{\Gamma(\beta k+1)}
=x_0E_\beta((\mu+\sigma^2/2)t^\beta),
$$
where $E_\beta(z)=\displaystyle\sum_{k=0}^{\infty}z^k/\Gamma(\beta k+1)$ 
is the Mittag-Leffler function of order $\beta$ \cite{mamitt}. Similarly:
$$
E(S(l_\beta(t))^2)=x_0^2E\left(\exp\left[(2\mu+3\sigma^2)l_\beta(t)\right]\right)=
x_0^2E_\beta((2\mu+3\sigma^2)t^\beta).
$$
Finally one has:
\begin{equation}
\left\{
\begin{array}{ll}
m(t)=E(D(t))=x_0E_\beta((\mu+\sigma^2/2)t^\beta)\\[0.5cm]
\sigma^2(t)=E(D(t)^2)-m(t)^2=x_0^2\left[E_\beta((2\mu+3\sigma^2)t^\beta)-E_\beta((\mu+\sigma^2/2)t^\beta)^2\right]
\end{array}
\right.
\label{geommom}
\end{equation}

\subsubsection{Exponential-decay kernel}
We now consider the {\it exponential-decay kernel} $K(t)=e^{-at}$, $a\ge 0$. 
The non-Markovian Fokker-Planck equation is:
\begin{equation}
u(x,t)=u_0(x)+\int_{0}^{t} e^{-a(t-s)}\left[(2\sigma^2-\mu)+(4\sigma^2-\mu)x\partial_x+\sigma^2 x^2 \partial_{xx}\right]u(x,s)ds,\;\;a\ge 0.
\label{GeomFlexp}
\end{equation}
We denote by $\mathcal{G}(x,t)$ the fundamental solution of the Markovian equation; 
namely Eq. (\ref{logn})
\begin{equation}
\mathcal{G}(x,t)=\frac{1}{x|\sigma|\sqrt{4\pi t}}\exp\left(-\frac{\Big{(}\log(x/x_0)-(\mu-\sigma^2/2)t\Big{)}^2}{\sigma^24t}\right),\;\; x,t\ge 0.
\end{equation}
Then, using Eq. (\ref{exptime}), the fundamental solution of Eq. (\ref{DrfFlexp}) is:
\begin{equation}
f(x,t)=e^{-at}\mathcal{G}(x,t)+(1-e^{-at})\Phi(x,t),
\end{equation} 
where:
$$
\Phi(x,t)=
\frac{1}{1-e^{-at}}\left[\frac{a}{|\sigma| x}e^{\log(\frac{x}{x_0})
(\frac{2\mu-\sigma^2}{4\sigma^2})}\int_0^tG(x',\tau)e^{- a'  \tau}\,d\tau
			\right],\;\; x\ge 0
$$
and:
\begin{equation}
\left\{
\begin{array}{ll}
a'=\frac{(\mu-\sigma^2/2)^2}{4\sigma^2}+4a, & a,\mu\ge 0,\;\; \sigma>0,\\[0.3cm]
x'=\log(x/x_0)/\sigma, & x\ge 0.
\end{array}
\right.
\end{equation}
Thus, as in Eq. (\ref{fonexp}), we have:
\begin{prop}
$$
f(x,t)=e^{-at}\mathcal{G}(x,t)
$$
$$
+\frac{a}{|\sigma| x}e^{\log(\frac{x}{x_0})(\frac{2\mu-\sigma^2}{4\sigma^2})}
\left\{\frac{1}{4\sqrt{a'}}\exp(x'\sqrt {a'})\textrm{\rm Erf}\left(\frac{x'}{2\sqrt t}+\sqrt{a't}\right)-
\exp(-x'\sqrt {a'})\textrm{\rm Erf}\left(\frac{x'}{2\sqrt t}-\sqrt{a't}\right)  \right\}
$$
\begin{equation}
-\frac{a}{4|\sigma| x\sqrt{a'}}e^{\log(\frac{x}{x_0})
(\frac{2\mu-\sigma^2}{4\sigma^2})}\sinh (|x'|\sqrt{a'}).
\label{fonexpgeom}
\end{equation}
\end{prop}
The stationary distribution, obtained as $t\rightarrow \infty$, is:
\begin{equation}
\overline \Phi(x)=
\frac{a}{4|\sigma| x\sqrt{a'}}\exp\left(\log\Big{(}\frac{x}{x_0}\Big{)}\Big{(}\frac{2\mu-\sigma^2}{4\sigma^2}\Big{)}\right)(\cosh(x'\sqrt{a'})-\sinh (|x'|\sqrt{a'}).
\end{equation}

\section{Conclusions}\label{C}
Theorem  \ref{t2} states that the fundamental solution $f(x,t)$ 
of a non-Markovian diffusion equation of the form (\ref{nmeq2})
\begin{equation}
u(x,t)=u_0(t)+\int_{0}^{t}g'(s)K\left(g(t)-g(s)\right)\mathcal{P}_xu(x,s)ds,\;\; x\in \mathbb{R},\;\; t\ge 0,
\label{c1}
\end{equation}
is 
\begin{equation}
f(x,t)=\int_{0}^{\infty}\mathcal{G}(x,\tau)h(\tau,g(t))d\tau,
\label{c3}
\end{equation}
where $\mathcal{G}(x,t)$ is the fundamental solution of the Markovian equation (\ref{fPl}) 
and $h(\tau,t)$ is the fundamental solution of the non-Markovian forward drift equation 
\begin{equation}
u(\tau,t)=u_0(\tau)-\int_{0}^{t}K(t-s)\partial_\tau u(\tau,s)ds,\;\; \tau,t\ge 0,
\label{c2}
\end{equation}

If the memory kernel $K(t)$ is chosen in a suitable way (see Section \ref{St}), 
the solution $f(\cdot ,t)$ preserves non-negativity and normalization for all $t\ge 0$. 
Thus, it can be interpreted as the marginal density function of a non-Markovian stochastic process. 
In view of Eq. (\ref{c3}), this stochastic process is naturally interpreted 
as a subordinated process Eq. (\ref{sQ}).\\

We focused on two kind of memory kernels: 
the power kernel $K(t)=t^{\beta-1}/\Gamma(\beta)$, $0<\beta\le 1$, 
and the exponential-decay kernel $K(t)=e^{-at}$, $a\ge 0$. 
 
The first provides the so-called time-fractional Fokker-Planck equations  (\ref{tfpll}). 
In particular we studied the case $\mathcal{P}_x=\partial_{xx}$ 
(see Section \ref{sstfde}), which corresponds to the choice of a 
``standard'' Brownian motion for the parent Markov model. 
In this case, the fundamental solution can be written in terms of 
an entire transcendental function, see Eq. (\ref{gbm}), 
and is related to a  Fox $H$-function through Eq. (\ref{Mfox}). 
We have also considered more complicated cases, 
namely Brownian motion with drift $\mu$ (see Section \ref{bmwdr}) 
and Geometric Brownian motion (see Section \ref{gmbrw}). 
In these cases the fundamental solutions can be written in terms of a 
superposition of Fox $H$-functions, see Eq. (\ref{phiH} and Eq. (\ref{phigH}). 

The exponential-decay kernel corresponds heuristically to 
a system in which the non-local memory effects are negligible 
for small times. 
In fact, the fundamental solution can always be written in the form of Eq. (\ref{brownber}),
$$
f(x,t)=e^{-at}\mathcal{G}(x,t)+(1-e^{-at})\phi(x,t),\;\; t\ge 0,
$$
where $\mathcal{G}(x,t)$ is the fundamental solution of the Markovian equation, 
and where the function $\phi(x,t)$ 
is a probability density which becomes stationary for large times. 
Therefore, it is always possible to find stochastic 
models that become stationary for large times and whose marginal density is given by Eq. (\ref{c3}).\\  

However, see Subsection \ref{thenot}, 
the stochastic representation is not unique, that is, 
there are many different stochastic processes whose marginal density is $f(x,t)$. 
For example,  consider the case where $\mathcal{P}_x=\partial_{xx}$ and $g(t)=t$. 
Then, $f(x,t)$ is the marginal density of $B(l(t))$, $t\ge 0$, 
where $B(t)$  is a``standard'' Brownian motion and where $l(t)$ 
is a random time process satisfying Eq. (\ref{c2}). 
If the random time $l(t)$ is required to be self-similar of order $\beta$,  
then in view of Theorem \ref{ppp}, the memory kernel must be a power function 
$K(t)=t^{\beta-1}/\Gamma(\beta)$ with $0<\beta\le 1$. 
The corresponding non-Markovian diffusion equation (\ref{c1}) 
is called in this case time-fractional diffusion equation of order $\beta$ 
(see Subsection \ref{sstfde}). 
The corresponding random time process $l(t)=l_\beta(t)$, 
can be the local time of a $d=2(1-\beta)$-dimensional fractional 
Bessel process or, alternatively, 
the inverse of the totally skewed strictly $\beta$-stable process. 
However, $f(x,t)$ is also the marginal density of the process 
$Y(t)=\sqrt {l_\beta(1)}B_{\beta/2}(t)$, 
where $B_{\beta/2}$ is a fractional Brownian motion independent of the random time $l_\beta(t)$. 
In all the previous examples, the self-similarity parameter  $H=\beta/2$ 
is restricted to the region $0<H\le 1/2$. 
We can obtain stochastic processes with higher values of the self-similarity parameter 
by introducing the time-scaling function $g(t)$. 
In this way, for example choosing $g(t)=t^{\alpha/\beta}$, $0<\alpha<2$, 
we obtain the process $D(t)=B(l_\beta(t^{\alpha/\beta}))$, $t\ge 0$, 
and the process $\mathcal{Y}(t)=\sqrt {l_\beta(1)}B_{\alpha/2}(t)$, $t\ge 0$, 
which are self-similar with parameter $H=\alpha/2$ so that $0<H<1$ (see Subsection \ref{sttfra}). 
In contrast to $D(t)$ the process $\mathcal{Y}(t)$ has stationary increments.\\  

The solution of the ``non-Markovian'' equation (\ref{c1}) 
can be stated explicitly in all the cases considered. 
We computed it analytically and graphed it in particular cases. 
This solution is a marginal (one-point) density function. 
We have then presented various random processes whose marginal density function coincides with that solution.

\section*{Acknowledgments}
This work has been carried out in the framework of a research 
project for {\it Fractional Calculus Modelling}
(URL: {\tt www.fracalmo.org}).
It was pursued while Antonio Mura was visiting Boston University 
as a recipient of a fellowship of the Marco Polo project of the
 University of Bologna. 
The authors appreciate partial support by 
the NSF Grants DMS-050547 and DMS-0706786 at Boston University,
 by  the Italian Ministry of University (M.I.U.R) through
 the Research Commission of the
 University of Bologna, and by the National Institute of Nuclear
Physics (INFN)	through the Bologna branch (Theoretical Group).
Finally, the authors would like to thank the  anonymous referees for
their comments.



\end{document}